\documentclass[prd,twocolumn,showpacs,nofootinbib,preprintnumbers]{revtex4}

\usepackage{amsmath}
\usepackage{amsfonts}
\usepackage{graphicx}
\usepackage[T1]{fontenc}
\usepackage{hyperref}

\def\be{\begin{equation}}
\def\ee{\end{equation}}
\def\ba{\begin{eqnarray}}
\def\ea{\end{eqnarray}}
\def\bs{\begin{subequations}}
\def\es{\end{subequations}}
\newcommand{\vp}{\varphi}
\newcommand{\ve}{\varepsilon}
\newcommand{\wteta}{\widetilde{\theta}}
\newcommand{\dth}{\bar{\theta}}
\newcommand{\dq}{\bar{q}}
\newcommand{\dy}{\bar{y}}
\newcommand{\da}{\bar{a}}
\newcommand{\de}{\bar{\epsilon}}
\newcommand{\dH}{\bar{H}}
\newcommand{\dalp}{\bar{\alpha}}

\begin{document}

\title{Patch dualities and remarks on nonstandard cosmologies}
\author{Gianluca Calcagni}
\email{calcagni@fis.unipr.it}
\affiliation{Dipartimento di Fisica, Universit\`{a} di Parma\\}
\affiliation{INFN -- Gruppo collegato di Parma\\ Parco Area delle Scienze 7/A, I-43100 Parma, Italy}
\date{October 21, 2004}
\begin{abstract}
In this paper we establish dualities between inflationary, cyclic/ekpyrotic, and phantom cosmologies within the patch formalism approximating high-energy effects in scenarios with extra dimensions. The exact dualities relating the four-dimensional spectra are broken in favour of their braneworld counterparts; the dual solutions display new interesting features because of the modification of the effective Friedmann equation on the brane. We then address some qualitative issues about phantomlike cosmologies without phantom matter.
\end{abstract}
\preprint{gr-qc/0410027}
\maketitle


\section{Introduction}

The ultimate theory of everything, if any, is a long-living mirage that physicists and mathematicians have been pursuing for years in the attempt to solve many fundamental problems rooted in our modern view of the Universe. One of the open issues is how to reconcile general relativity and quantum physics, two separate branches that experiments and observations has widely accepted as meaningful descriptions of natural phenomena, at least each in its own range of influence. The marriage between the two would require deep modifications of both and, although great progress has been made in this direction thanks to string theory, a happy ending to the story is still missing. In particular, the  mostly successful big bang model of cosmological evolution, which manages to glue gravitation and microphysics together in a very nontrivial way, sits on the paradox of the initial singularity: the original point from which all came decrees the failure of general relativity as a self-contained framework, since the relevant cosmological quantities diverge by definition when going back to the first instant of the past. At present we know that quantum effects can resolve such a point into a finite speck and smooth out the worried infinities \cite{strin}.

From a philosophical perspective, the big bang has raised many questions about the nature of time and its birth, leading to the (indeed not new) hypothesis that the Universe may experience a cyclic succession of expansions and contractions in which the big bang singularity is just a transitory phase (a bounce) in a wider process of evolution; see \cite{old} for old attempts to implement this idea. At a semiclassical level, the structure of the perturbations generated through the bounce can be more complicated than the standard one in a monotonically expanding universe; for example, vector modes cannot be neglected during the contracting phase in contrast to their decaying behaviour in the post big bang phase \cite{BB}. General phenomenology of cyclic models and bouncing cosmological perturbations have been studied, e.g., in \cite{PP1,CDC,PPG,KST,EWST,AW}; the case of a bouncing closed universe has been investigated in \cite{close}. 

Moreover, bouncing flat cosmologies may require a violation of the null energy condition $\rho+p\geq 0$, where $\rho$ and $p$ are the energy density and pressure of a perfect fluid describing the matter content of the early Universe \cite{BM,PP1,AW,PP2,PPG}. The field associated to an equation of state $p=w\rho$ with $w<-1$ is called ``phantom'' \cite{pha1} and its nonconventional properties, including superacceleration ($\ddot{a}/a>H^2$), can give rise to a new sort of singularity (the big rip) as well as to an explanation of current observations of dark energy. For a reference list on the subject, see \cite{cal3} to which we add \cite{ACL,phant}.

An interesting singularity-free setup, alternative to inflation and motivated by string theory, is the ekpyrotic scenario, which explains the large-scale small anisotropies of the cosmic microwave background via a collision between wrinkling branes \cite{ekpy1,KOSST,ekpy2}. A general-relativistic treatment of ekpyrotic/cyclic scenarios predicts a scale-invariant scalar spectrum (with scalar index $n_s-1\approx 0$) and a blue-tilted tensor spectrum $n_t\approx 2$, while standard inflation generates almost scale-invariant spectra. In the latter case, this is a consequence of the slow-roll (SR) approximation, stating that both the parameter
\be \label{epsilon}
\epsilon \equiv -\frac{d \ln H}{d \ln a}=-\frac{\dot{H}}{H^2}\,,
\ee
and its time derivative must be sufficiently small.\footnote{In the following we will refer to $\epsilon$ as the ``SR parameter'' even when the slow-roll approximation $\epsilon \ll 1$ is not applied.} Conversely, the cyclic model achieves scale invariance when $\bar{\epsilon}\equiv \epsilon_\text{cyclic} \gg 1$. Recently, two remarkable dualities were discovered in flat cosmology, one relating inflationary to ekpyrotic/cyclic spectra \cite{GKST,dual,lid04} and the other connecting inflationary to phantom spectra  \cite{ACL,phdu,lid04}. More precisely, given an inflationary model there exist both cyclic and phantom cosmologies with the same spectra and such that
\bs\label{4Dduality}\ba 
\bar{\epsilon} &=& 1/\epsilon\,,\label{4Ddual}\\
\hat{\epsilon} &=& -\epsilon\,,\label{4Dphdual}\\
\bar{\epsilon} &=& -1/\hat{\epsilon}\,,
\ea\es
where $\hat{\epsilon}\equiv \epsilon_\text{phantom}$. In four dimensions, this triality is exact for arbitrary (even varying) $\epsilon$ \cite{lid04}. Other dualities can be found in \cite{altdu}.

The search of viable bouncing mechanisms has led to explore several possibilities that involve, for instance, varying couplings \cite{BKM}, noncommutative geometry \cite{MMMZ}, quantum gravity and cosmology \cite{quang} (see also \cite{sri04}). In particular, a Randall-Sundrum (RS) modification of the Friedmann equations has been considered \cite{SS}, in which a phantom component may help to tear apart black holes during the bounce \cite{BFK,PZ}. 

In braneworld scenarios the visible universe is confined into a (3+1)-dimensional variety (a brane) embedded in a larger noncompact spacetime (the bulk). One of the first problems one has to deal with when constructing such models is how to stabilize the extra dimension. This can be achieved in a number of ways; in the RS example, Goldberger and Wise have provided a mechanism according to which a 5D massive scalar is put into the bulk with a potential of the same order of the brane tension $\lambda$ \cite{GW}. If the energy density $\rho$ on the brane is smaller than the characteristic energy of the scalar potential, $\rho/V \sim \rho/\lambda \ll 1$, then the radion is stabilized and one gets the standard Friedmann equation $H^2 \propto \rho$ on the brane. On the contrary, if the brane energy density is comparable with the stabilization potential, $\rho/\lambda \gtrsim 1$, the bulk backreacts because it feels the presence of the brane matter, the minimum of the potential is shifted, and the well-known quadratic corrections to the Friedmann equation arise \cite{quco}.

Here we shall follow the second alternative and consider nonstandard cosmological evolutions on the brane, extending the RS discussion to arbitrary scenarios we dubbed ``patch cosmologies'' \cite{cal3}, with
\be \label{FRW}
H^2=\beta_q^2 \rho^q\,.
\ee
Here $H$ is the effective Hubble rate experienced by an observer on the brane, $\beta_q$ is a constant we will assume to be positive, and the exponent $q$ is equal to 1 in the pure 4D (radion-stabilized) regime, $q=2$ in the high-energy limit of the RS braneworld, and $q=2/3$ in the high-energy limit of the Gauss-Bonnet (GB) scenario;\footnote{Gauss-Bonnet cosmologies have been studied, e.g., in \cite{cal3,gabo} and references therein.} if braneworld corrections are important in the early Universe, one can follow the cosmological evolution through each energy patch in a given time interval where the patch approximation is valid.
	 					     
In this paper (Secs. \ref{setup} and \ref{patdua}) we will investigate the above-mentioned triality for general $q$ and show that Eq. (\ref{4Dduality}) no longer realizes exact correspondences between cyclic, inflationary, and phantom patches. According to the new dualities we will establish, any expanding universe is mapped to either a contracting or phantom universe which no longer display exactly the same scalar perturbations; our results are in agreement with previous investigations \cite{lid04}. Since braneworld spectra are broken under duality, and mapped into a quantitatively different contractinglike or phantomlike spectra, these transformations are not symmetries in the strict meaning of the word.

We will define the dual transformations both in a given patch by imposing $\bar{q}=q$ and between different patches, which we will call cross dualities. In addition, it will turn out that the generalized version of the 4D contracting (phantom) mapping gives rise to a phantom (contracting) dual solution when flipping the sign of $q$.

At last, inspired by a modified version of the phantom duality we shall outline some proposals for (i) a new bouncing scenario, (ii) the generation of features in the power spectrum breaking scale invariance, and (iii) an alternative to standard inflation. A full detailed implementation of these topics in a rigorous theoretical framework is beyond the scope of this article and will be left for the future; rather, we shall provide some preliminary comments in Sec. \ref{qbounce}. Conclusions are in Sec. \ref{concl}.


\section{Setup and preliminary remarks} \label{setup}

We assume there is a confinement mechanism on the brane for a perfect fluid with continuity equation $\dot{\rho}+3H\rho(1+w)=0$, and neglect any brane-bulk exchange. For an homogeneous scalar field $\phi$ with potential $V$,
\be
\rho(\phi)=\frac{\ell}{2}\dot{\phi}^2+V(\phi)\,,
\ee
while for a Dirac-Born-Infeld (DBI) tachyon
\be
\rho(T) = \frac{V(T)}{\sqrt{1-\ell\dot{T}^2}}\,.
\ee
Here $\ell=1$ for ordinary causal fields and $\ell=-1$ for phantoms. We will use the symbol $\vp$ to indicate the inflaton field in expressions valid for both the normal scalar and the tachyon.\footnote{To avoid confusion, we will call ``standard (phantom) ordinary scalar'' the $\phi$ field with $\ell=1$ ($\ell=-1$), ``standard (phantom) tachyon'' the $T$ field with $\ell=1$ ($\ell=-1$) and ``scalar'' (or, sometimes, ``inflaton'') the field satisfying the continuity equation, regardless of its action.} Equation (\ref{epsilon}) defines the time variation of the Hubble radius $R_H\equiv H^{-1}$; this parameter can also be expressed either via derivatives of the inflaton field, indicated with primes, or through the continuity and Friedmann equations:
\ba 
\epsilon &=& -\frac{a}{a'}\frac{H'}{H}\label{epsi1}\\
				 &=& \frac{3q(1+w)}{2}-\frac{\dot{q}}{q}\frac{\ln H^2}{2H}\,.\label{epsi2}
\ea
In the last formula, we have set $\beta_q=1$ and considered the general case of time-dependent $q(t)$. When $q=\text{const}$ (which we shall assume throughout this section), then $\epsilon>0$ when $\text{sgn}(q)=\text{sgn}(w+1)$, while phantom matter with $q>0$ (or ordinary matter with $q<0$) reverses the sign of $\epsilon$.

We can see that the dualities (\ref{4Ddual}) and (\ref{4Dphdual}) are broken in their simplest form when considering nontrivial patches. A first evidence comes from the equations of motion of scalar perturbations, which in four dimensions are invariant under the mapping $\epsilon \rightarrow \epsilon^{-1}$ for dominant and subdominant modes, separately \cite{GKST,dual}. Let $\Phi$ be the Newtonian potential and ${\cal R}$ the curvature perturbation related to the scalar spectrum generated by quantum fluctuations of a standard ordinary scalar field $\phi(x)=\phi(\tau)+\delta\phi(\mathbf{x},\tau)$ around a homogeneous background, where $\tau$ is conformal time. Defining $z\equiv\vartheta^{-1}\equiv a\dot{\phi}/H$ and the two Mukhanov variables $v\equiv-\Phi/\dot{\phi}$ and $u\equiv-z{\cal R}$ [with a $(+,-,-,-)$ metric; notations may vary according to the paper], the effective 4D equations of motion for the (Fourier-transformed) scalar-perturbation modes in the longitudinal gauge are (e.g., \cite{MFB})
\bs\label{mukha}\ba
\left(\partial_\tau^2+k^2-\frac{\partial_\tau^2\vartheta}{\vartheta}\right)v_k &=& 0\,,\\
\left(\partial_\tau^2+k^2-\frac{\partial_\tau^2 z}{z}\right)u_k &=& 0\,.
\ea\es
When $q\neq 1$ and the Friedmann equation receives corrections from the extra-dimensional physics, several arguments show that the Mukhanov equations (\ref{mukha}) still hold, at least at lowest SR order (quasi-de Sitter regime) and under the assumption of negligible Weyl contribution. We can express $\vartheta$ in terms of the slow-roll parameter (\ref{epsilon}) and its variation $\gamma = d\ln \epsilon/d{\cal N}$ with respect to 
\be
{\cal N} \equiv \ln\frac{a_fH_f}{aH}\,,
\ee
where the subscript $f$ denotes evaluation at the end of the inflationary or ekpyrotic phase; the standard ``forward definition'' of the number of e-foldings, $N=\ln (a/a_i)$, is related to this quantity by $d{\cal N}=(\epsilon-1)dN$.\footnote{In \cite{cal3} the ``backward definition'' $N=\ln (a_f/a)$ is used instead.} The conformal time satisfies the relation
\be \label{confor}
\tau=\int \frac{d t}{a} = \frac{1}{(\epsilon-1)aH}\,;
\ee
neglecting $O(\gamma^2)$ and $O(d\gamma/d{\cal N})$ terms, one finds
\ba
\frac{\partial_\tau^2\vartheta}{\vartheta}\tau^2 &\approx& \left(1+\frac{\theta}{2}\right) \left(1+\frac{\theta}{2}\epsilon\right)\frac{\epsilon}{(\epsilon-1)^2}+\frac{\epsilon^2-1}{2(\epsilon-1)^2}\gamma\,,\nonumber\\\label{exam}
\ea
where $\theta\equiv 2(1-q^{-1})$ and is equal to 0 in the 4D case, 1 in RS, and $-1$ in GB. In general relativity, a first step towards the duality $\epsilon \leftrightarrow \epsilon^{-1}$ is to note that the equation of motion for $v$ is invariant under the mapping (\ref{4Ddual}). However, when $\theta \neq 0$ this duality is explicitly broken by the term inside the second round brackets, which by the way contributes to the only piece surviving for a constant $\epsilon$ ($\gamma=0$). In the case of a standard tachyonic field, Eq. (\ref{exam}) has an extra term proportional to $\gamma(3+\theta\epsilon)(\epsilon-1)\epsilon/(3q-2\epsilon)$, which breaks the invariance even in four dimensions.


\section{Patch dualities} \label{patdua}

\begin{figure}
\includegraphics[width=8.6cm]{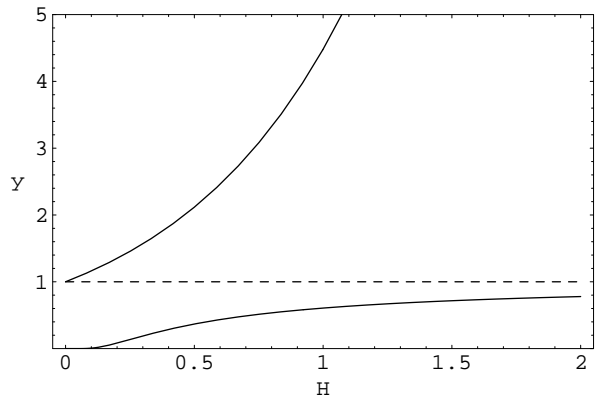}
\caption{\label{fig1}
The function $y(H)$ in the GB (upper curve) and RS (lower curve) expanding braneworlds. The image of $H$ is $y\geq 1$ in the first case and $0\leq y< 1$ in the second one.}
\end{figure}

We can make the previous argument more rigorous by means of the Hamilton-Jacobi formulation of the cosmological dynamics. Let us write down the Hamilton-Jacobi equations for the scalar field \cite{cal3},
\ba \label{hjphi}
V(\phi) &=& \left(1-\frac{\epsilon}{3q}\right)|H|^{2-\theta}\,,\label{Vphi}\\
V^2(T) &=& \left(1-\frac{2\epsilon}{3q}\right)H^{2(2-\theta)}\,,\label{Vtac}\\
H'a' &=& -\frac{3q}{2}\ell |H|^{\wteta}Ha\,,\label{hj}
\ea
where $\wteta=\theta$ for the ordinary scalar and $\wteta=2$ for the tachyon and we have set $\beta_q=1$. The absolute value of $H$ is necessary and sufficient to preserve the invariance under time reversal of the original equations of motion. Define the variable $y(\vp)$ as
\ba 
y(\phi)&\equiv& H(\phi)^{2\ell/3}\,,\qquad\qquad\qquad \theta=0\,,\label{4Dy}\\
y(\vp) &\equiv& \exp \left[\alpha |H(\vp)|^{-\wteta}\right]\,,\qquad \wteta \neq 0\,,\label{y}
\ea
where the coefficient $\alpha \equiv -2\ell/(3q\wteta)$ is $\alpha=-1/3$ and $\alpha=1$ for a RS and GB braneworld without phantoms, respectively. Then, $\text{sgn}(dy)=\text{sgn}(dH)$ when $q>0$, and Eq. (\ref{hj}) can be recast as
\be
y(\vp)'a(\vp)'=- y(\vp)a(\vp)\,.\label{hj2}
\ee
Figure \ref{fig1} shows the function $y(H)$ for the RS and GB case.
From now on we will set $\wteta=\theta$ for lighter notation. The parameter $\epsilon$ can be written as
\be
\epsilon = -\frac{(\ln y)'^2}{\theta \ln y}\,.
\ee
The Hamilton-Jacobi equations encode all the dynamical information for the cosmological evolution. If two different models ($\vp$,$\theta$) and ($\vp'$,$\theta'$) display the same set of equations, then we will say there is a duality between them. Let us now consider what transformations are symmetries of Eq. (\ref{hj2}). In general, a symmetry transformation can be written as
\bs\label{map1}\ba
\da(\vp)&=&f_1(\vp)\,,\\
\dy(\vp)&=&f_2(\vp)\,,\label{map1b}
\ea\es
provided that $[\ln f_1(\vp)]'[\ln f_2(\vp)]'=-1$. In Eq. (\ref{map1b}) all the elements of $\dy$, including $\theta$, are evaluated in the dual patch. Since in principle it is not possible to set $\beta=1=\bar{\beta}$ consistently, one should restore the dimensional factors in the previous and following expressions, noting that $[\beta]=E^{(\theta+2)/(\theta-2)}$.


\subsection{Singular dualities}

A simple realization of Eq. (\ref{map1}) is
\bs\ba
\da(\vp)&=&y(\vp)^{p(\vp)}\,,\\
\dy(\vp)&=&a(\vp)^{1/p(\vp)}\,.
\ea\es
In order to satisfy the above integrability condition, the function $p(\vp)$ must be either a constant or
\be \label{p}
p(\vp)=p_0\frac{\ln a(\vp)}{\ln y(\vp)}\,,
\ee
where $p_0$ is an arbitrary real constant. For constant $p=p_0$,
\bs\label{map2}\ba
\da(\vp)&=&y(\vp)^{p_0}\,,\label{map2a}\\
\dy(\vp)&=&a(\vp)^{1/p_0}\,,
\ea\es
also considered in \cite{CLLM}. It is convenient to define the new parameter
\be\label{ve}
\ve \equiv \frac{\epsilon}{q|H|^{\theta}}= \frac{3}{2}\ell\left(\frac{a}{a'}\right)^2;
\ee
then one can express the spectral amplitudes as $A_s(\phi)^2\propto H^2/\ve$, $A_s(T)^2\propto H^\theta/\ve$, and get the spectral indices from the evolution equation $\dot{\ve}=2H\ve(\epsilon-\eta)$, which reproduces the 4D one when $\ve=\epsilon$. The set of equations describing the dual solution can be obtained from Eqs. (\ref{map2a}), (\ref{y}), and (\ref{ve}):
\ba
\da(\vp) &=& \exp \left(-p_0\int^\vp d\vp \frac{a}{a'}\right),\\
|\dH(\vp)| &=& \left[\frac{\dalp p_0}{\ln a(\vp)}\right]^{1/\dth},\label{Hdual}\\
\bar{\ve}(\vp)\,\ve(\vp) &=& \frac{9\ell\bar{\ell}}{4p_0^2}\,.\label{epsidual}
\ea
The right-hand side of Eq. (\ref{Hdual}) is positive when $\text{sgn}(\theta)=\text{sgn}(1-a)$ and $q>0$. Then $a<1$ for RS and tachyon scenarios and $a>1$ for the GB braneworld. Equation (\ref{epsidual}) reproduces Eq. (\ref{4Ddual}) in the ordinary scalar case with $\theta=0$ and $p_0=3/2$, although the 4D auxiliary variable $y$, Eq. (\ref{4Dy}), is constructed in a different way. In the case of cyclic duality ($\bar{\ell}=\ell=1$), the mapping (\ref{map2}) relates a standard  accelerating ($\epsilon<1$) expanding universe with a standard decelerating ($\bar{\epsilon}>1$) contracting phase with the typical properties of cyclic cosmology.

The transformation (\ref{map2}) connects the scale factor of the expanding cosmology to that of a dual cosmology when expressed in terms of the scalar field. In the dual model, the scalar field acquires a different time dependence relative to its expanding counterpart. The time variable can be written as an integral over $\vp$,
\be\label{tnorm}
t=\int^\vp \frac{d\vp}{H} \frac{a'}{a}\,;
\ee
the time variable $\bar{t}$ of the dual solution is then
\be
\bar{t} = \frac{2\ell p_0}{3}\int^\vp \frac{d\vp}{a^{3\bar{\ell}/(2p_0)}} \frac{H'}{H}\,,\label{4Dt}
\ee
in the 4D$\to$4D case,
\be
\bar{t} = \frac{2\ell p_0}{3}\int^\vp d\vp\, (\ln a)^{1/\dth}(\ln H)'\,,
\ee
for the pure braneworld dual of the 4D scenario, while for a general cross duality with $\dth\neq0\neq\theta$, using Eqs. (\ref{y}) and (\ref{hj2}),
\be
\bar{t} = -\frac{p_0}{(\dalp p_0)^{1/\dth}}\int^\vp d\vp \frac{(\ln a)^{1/\dth}}{(\ln a)'}\,.\label{t}
\ee
Everywhere we have omitted $\text{sgn}(\bar{H})$ which is implicit in the time-reversal symmetry of the dual solution.
The dual evolution of the scalar field will be denoted as $\bar{\vp}(t)\equiv \vp(\bar{t})$. For $\bar{q}=q=1$ and $p_0=3/2$, these relations reproduce the already known four-dimensional standard triality.

The exact inversion of the SR parameter $\bar{\epsilon}\epsilon=1$ is achieved in any dimension by the stationary cosmology $a(t)=t$. Otherwise, the fixed points of the transformation (\ref{map2}) are those with 
\be \label{self}
\ve_\text{self-dual} \equiv \frac{3}{2p_0}\,.
\ee
In general, we define a \emph{self-dual solution} as the set of roots of Eq. (\ref{self}). In four dimensions with $p_0=3/2$, Eq. (\ref{self}) reduces to the self-dual condition $\epsilon=1$. From the dual of the SR parameter as given by Eq. (\ref{ve}), it is clear that dual cosmologies superaccelerate either in the phantom case with $\dq>0$ or in the ordinary one for $\dq<0$.

Let us discuss what is the structure of the cyclic duality in a patch framework with positive $q$ and $p_0$. For clarity, we compare the cases $\theta=0,\pm 1$. By definition, standard inflation is characterized by a monotonically varying scalar field which can be assumed to be increasing with time, $\dot{\vp}=Ha/a'>0$. A parity transformation $\vp\to-\vp$ always achieves this condition. Therefore $a'>0$ (since $H>0$) and $H'<0$. On the contrary, the dual scale factor is a decreasing function of $\bar{\vp}$ since $\bar{a}'/\bar{a}\propto -a/a'<0$. 

In the four-dimensional ordinary scalar case, the expanding ($\bar{H}>0$) dual solution has $\dot{\bar{\phi}}=aH/H'<0$, $\bar{a}'<0$, and $\bar{H}'>0$; also, from Eq. (\ref{epsi1}) $\bar{\epsilon}>0$. Under the time reversal
\begin{eqnarray*}
t \in [0,+\infty[  &\to& t \in\,\,]-\!\!\infty,0]\,,\\
\dot{\bar{\vp}}(t)&\to& -\dot{\bar{\vp}}(-t)\,,\\
\bar{a}(t)        &\to& \bar{a}(-t)\,,\\
\bar{H}(t)        &\to& -\bar{H}(-t)\,,\\
\bar{\epsilon}(t) &\to& \bar{\epsilon}(-t)\,,
\end{eqnarray*}
the dual cosmology becomes contracting while keeping the condition $\dot{\bar{\phi}}>0$ and $\bar{\epsilon}>0$ (i.e., it does not superaccelerate). In a general expanding patch, the dual time evolution of the scalar field is $\dot{\bar{\vp}} \propto (\ln a)^{-1/\theta}$, which shows that in the RS, GB, and tachyon scenarios the evolution of the dual cosmology is not regular because of the factor $\ln a$.

To be consistent with the image of $y$ and Eqs. (\ref{map2}) and (\ref{Hdual}), we require $a<1$ in the RS scenario and $a>1$ in the GB one. In this case the above considerations hold with the same signs as in 4D and we get contracting solutions after a time reversal. Let $t_*$ be the time when $a(t_*)=1$; then $\infty>H>H_*=H(t_*)$ and the dual RS scale factor $\bar{a}$ ranges from $\bar{a}_*=\exp [-1/(2H_*)]$ to 1. In the GB case, $\infty>H_*>H>0$ and $\infty>\bar{a}>\exp(3H_*/2)$. As a matter of fact, in the example below the range of the GB power-law dual solution is modified according to the sign of the scalar field (negative for an expanding cosmology) but the underlying message in unchanged: Because of the different range of the variables involved in the mapping (\ref{map2}), the dual cosmology is only a portion of a contracting cosmology evolving from the infinite past to the origin. For this reason one might consider Eq. (\ref{map2}) as an ``incomplete'' mapping; rather, the restriction on the range of $a$ makes these solutions ``complete'' although very peculiar, since the dual Hubble parameter indeed goes from infinity to zero but in a finite time interval.

Note that these features are \emph{not} an effect of the patch approximation we have used for simplifying the cosmological evolution. For $\theta=1$, Eq. (\ref{y}) is a good approximation of the exact Randall-Sundrum case, where \cite{lid04}
\be \label{yRS}
y^2_\text{RS} \equiv \frac{\rho}{\rho+2\lambda}\,.
\ee
In order to make it manifest, we restore the dimensional factors and temporarily redefine the variable $y_\text{temp} \equiv y^{p_0}$ with $p_0=3/2$; then, in the patch approximation $\delta\equiv \lambda/\rho \ll 1$,
\be
y_\text{temp}= \exp \left(\frac{-\kappa_4^2}{6\beta_2H}\right) \approx 1-\sqrt{\frac{\lambda\kappa_4^2}{6H^2}}=1-\delta\,,
\ee
which reproduces Eq. (\ref{yRS}) in the high-energy RS limit. Note that even in the exact RS scenario $\epsilon$ is not exactly inverted under the transformation (\ref{map2}), since $\epsilon \propto [(1+y^2_\text{RS})/(1-y^2_\text{RS})](y'_\text{RS}/y_\text{RS})^2$. The dual Hubble parameter through $\da=y_\text{RS}$ is
\be\label{HRS}
\bar{H}(\phi)=\frac{\sqrt{2\lambda} a(\phi)}{1-a(\phi)^2}\,,
\ee
which is positive as far as $a<1$. It agrees with Eq. (\ref{Hdual}) in the above limit $a =\bar{y}_\text{temp}\approx 1-\bar{\delta}$, as $\bar{H}\approx (2\bar{\delta})^{-1}$. Also,
\be\label{vpRS}
\dot{\bar{\phi}}=-\ell\frac{\sqrt{8\lambda} a'}{1-a^2}\,.
\ee
Equations (\ref{yRS}), (\ref{HRS}), and (\ref{vpRS}) fully confirm what we have said about the structure of the dual solution, since the image of $y_\text{RS}$ is the same as that of $y$ for $q=2$.

\subsubsection{Self-dual solutions and power-law expansion: ordinary scalar case}

The self-dual solutions of the three scenarios with an ordinary scalar field are
\ba
a(t) &=& \exp\left[-p_0\exp \left(-t/p_0\right)\right]\,,\quad\qquad \theta=-1\,,\label{selfGB}\\
a(t) &=& \exp (\sqrt{2p_0t/3})\,,\qquad\qquad\qquad \theta=1\,,\label{selfRS}\\
a(t) &=& t^{2p_0/3} \,,\qquad\qquad\qquad\qquad\qquad \theta=0\,.\label{self4D}
\ea
As an example of the duality, let us consider the power-law inflation,
\be
a(t)=t^n\,,\qquad \epsilon = 1/n\,.
\ee
The ordinary scalar field associated with this expansion is such that $\dot{\phi}^2=2n^{1-\theta}/(3qt^{2-\theta})$. In four dimensions, the exact cosmological solution corresponding to this scale factor is
\be
\phi(t) = \phi_0 \ln t\,,\qquad V(\phi)=V_0e^{-2\phi/\phi_0}\,,
\ee
where $\phi_0=\sqrt{2n/3}$. The scale factor and Hubble parameter read
\be\label{pls}
a(\phi)=e^{n\phi/\phi_0}\,,\qquad H(\phi)=ne^{-\phi/\phi_0}\,,
\ee
respectively. From Eq. (\ref{4Dt}) with $p_0=3/2$, the (time reversed) cyclic-dual solution is
\be
\bar{a}(t)=(-t)^{1/n}\,,\qquad \bar{\epsilon} = n\,, \qquad \phi(t) =-\frac{2}{3\phi_0}\ln (-t)\,,
\ee
after a redefinition $n\bar{t}\rightarrow t$. The dual of the potential can be obtained by taking the dual of Eq. (\ref{Vphi}).

\begin{figure}
\includegraphics[width=8.6cm]{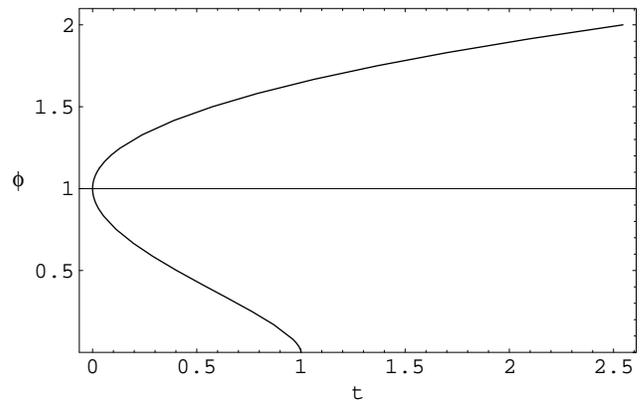}
\caption{\label{fig2}
The normalized RS scalar field $\phi$ as a function of time. The solid horizontal line divides the solutions of the duality (\ref{map2}) with $p_0=3/2$ (region $\phi<1$) and $p_0=-3/2$ (region $\phi>1$).}
\end{figure}

\begin{figure}
\includegraphics[width=8.6cm]{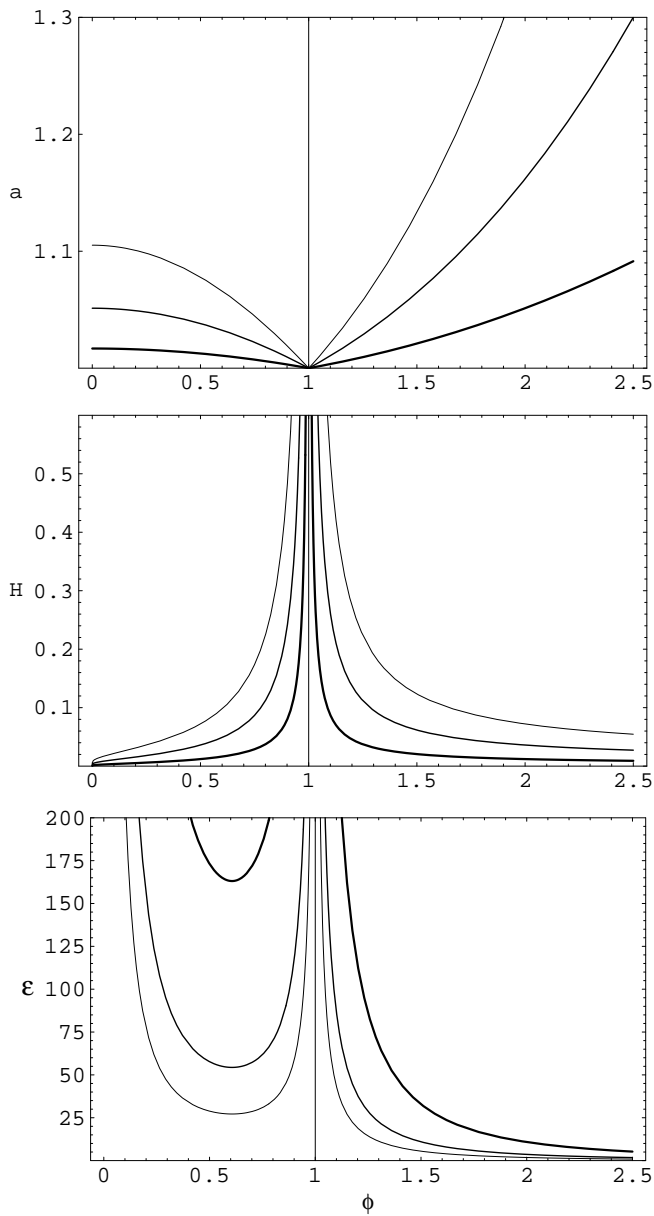}
\caption{\label{fig3}
The Randall-Sundrum solutions dual to RS power-law inflation for $n=5,10,30$ (increasing thickness). From top to bottom, each panel corresponds to the $\phi$ behaviour of the dual scale factor, the Hubble parameter and SR parameter under the duality (\ref{map2}) with $p_0=3/2$ (region $\phi<1$) and $p_0=-3/2$ (region $\phi>1$).}
\end{figure}

In the Randall-Sundrum scenario, the power-law expansion is realized by
\be
\phi(t)=\phi_0t^{1/2}\,,\qquad V(\phi)=V_0\phi^{-2}\,,
\ee
where $\phi_0=\sqrt{4/3}$. Then, after a redefinition $\phi/\phi_0 \to \phi$,
\be\label{rs}
a(\phi)=\phi^{2n}\,,\qquad H(\phi)=n\phi^{-2}\,,
\ee
and Eq. (\ref{t}) gives $t=1+\phi^2(\ln \phi^2-1)$; the dual RS cosmology under the mapping (\ref{map2}) has 
\bs\label{RSdual}\ba
\da &=& \exp[p_0(1-\phi^2)/(3n)]\,,\\
|\dH| &=& -p_0[3n\ln \phi^2]^{-1}\,,\\
\de&=& -3n(p_0\phi^2 \ln \phi^2)^{-1}\,,
\ea\es
where we have chosen the normalization of the scale factor such that $\da(1)=1$. A changing in the sign of $p_0$ results in different dual solutions. Figure \ref{fig2} shows the time behaviour of $\phi$ in the two separate regions $0<\phi<1$ and $\phi>1$; the quantities of Eq. (\ref{RSdual}) with $p_0=3/2$ are depicted in the left side ($\phi<1$) of Fig. \ref{fig3}, while the cyclic duals with $p_0=-3/2$ are in the right portion ($\phi>1$). Time flows from $\phi=1$, where the vertical line in each panel separates the two dual solutions. 

In the allowed region $\phi<1$ with $p_0>0$ (no phantoms, $\de>0$), the dual scale factor $\da(\phi)$ increases from $\da(1)$ to $\bar{a}(0)$ in a finite time interval, while the dual Hubble parameter goes from infinity to zero in the meanwhile. Solutions with $\phi>1$ and $p_0<0$ behave much better, since they extend not only up to the infinite future, but also are nonsingular at the origin, a very promising feature in classical bouncing models.

One gets the RS contracting solution simply by reversing the time direction [so that the dual scale factor $\bar{a}(\phi)$ decreases from $\bar{a}(0)$ or $\bar{a}(\infty)$ to $\bar{a}(1)$] and flipping the sign of $\bar{H}$. The dual slow-roll parameter does not change under time reversal and keeps being positive. 
By inverting Eq. (\ref{t}) in the region $\phi>1$, one gets the time dependence of the scale factor as $\bar{a}(t) \propto \exp[t/W(t/e)]$, where $W(x)$ is the product log function solving the nonlinear equation $x=W e^W$.

\begin{figure}
\includegraphics[width=8.6cm]{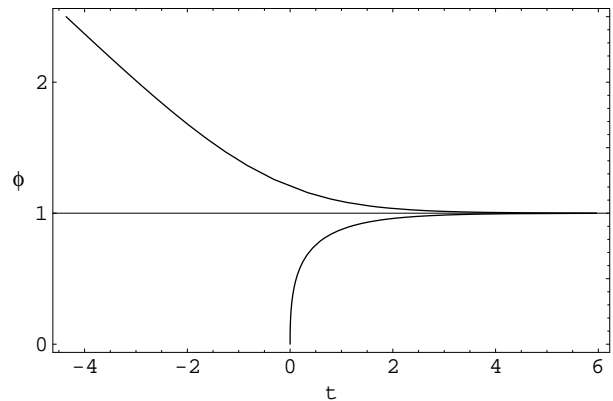}
\caption{\label{fig4}
The normalized GB scalar field $\phi$ as a function of time. The solid horizontal line divides the solutions of the duality (\ref{map2}) with $\text{sgn}(p_0)= \pm 1$ at the infinite future $\phi=1$.}
\end{figure}

\begin{figure}
\includegraphics[width=8.6cm]{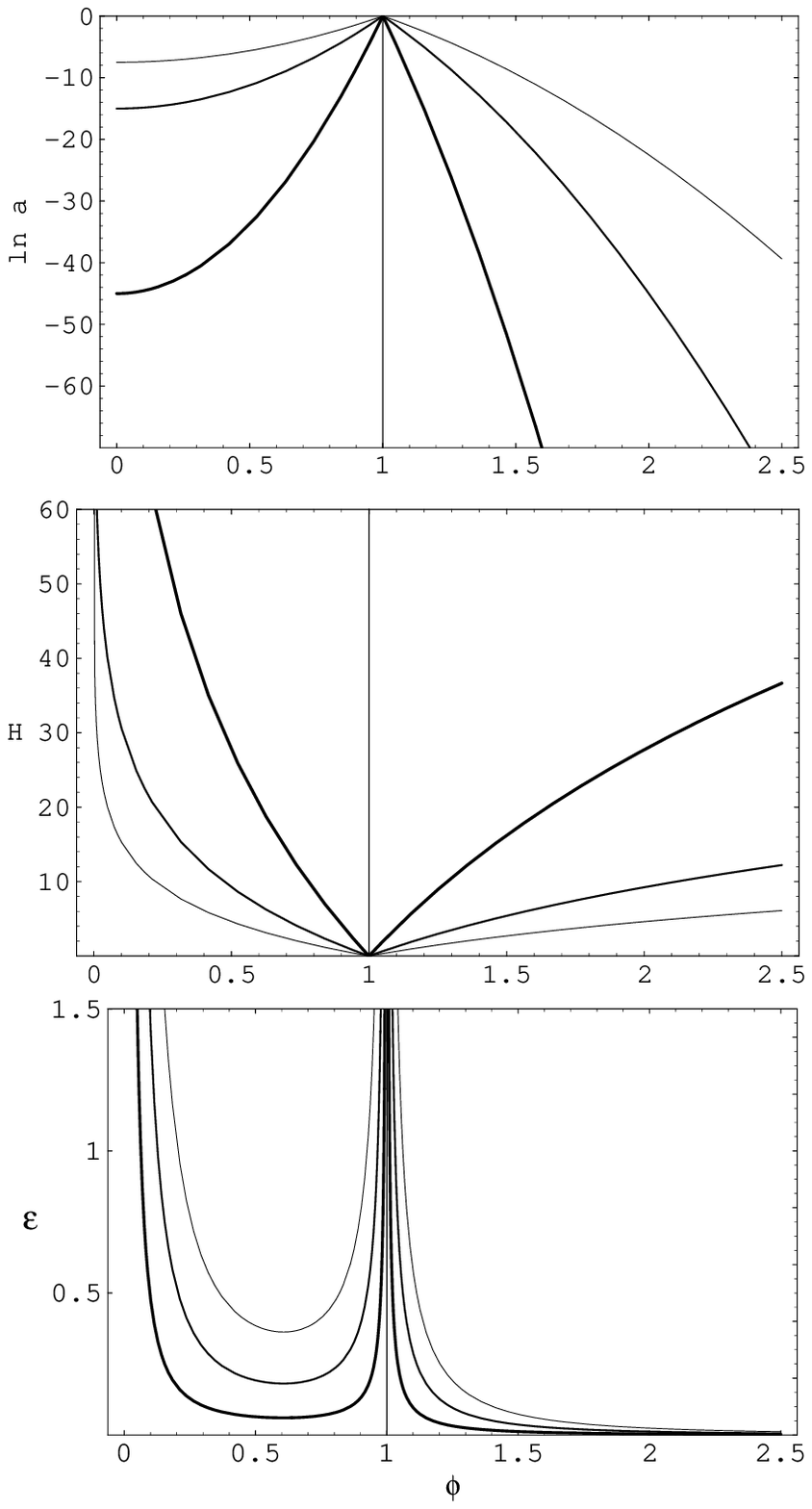}
\caption{\label{fig5}
The Gauss-Bonnet solutions dual to GB power-law inflation for $n=5,10,30$ (increasing thickness). From top to bottom, each panel corresponds to the $\phi$ behaviour of $\ln\bar{a}(\phi)$, $\bar{H}(\phi)$, and $\bar{\epsilon}(\phi)$ under the duality (\ref{map2}) with $p_0=3/2$ (region $\phi<1$) and $p_0=-3/2$ (region $\phi>1$).}
\end{figure}

In the Gauss-Bonnet case we have
\be
\phi(t)=-2nt^{-1/2}\,,\qquad V(\phi)=V_0\phi^6\,,
\ee
together with
\be
a(\phi) = \phi^{-2n}\,,\qquad H(\phi)=n\phi^2\,,
\ee
where $-\phi/2n\to \phi$. Equation (\ref{t}) gives $\bar{t}$ in terms of $\phi$: it turns out that $t =4\bar{t}/9=\int d\phi\, \phi (\ln \phi)^{-1}=-\text{Ei}[\ln \phi^2]$, where Ei is the exponential integral function plotted in Fig. \ref{fig4}. From Eqs. (\ref{map2a}), (\ref{Hdual}), and (\ref{epsidual}), the dual GB cosmology is
\bs\label{GBdual}\ba
\da &=& \exp[p_0 n(\phi^2-1)]\,,\\
|\dH| &=& -(n/p_0)\ln \phi^2\,,\\
\de &=& -(p_0n\phi^2\ln \phi^2)^{-1}\,,
\ea\es
and again the cyclic solution with ordinary matter evolves with $\da<\infty$ for all $t$ and $p_0>0$. On the contrary, in the branch with $p_0<0$ the dual scale factor $\da$ does not collapse to zero at the origin and diverges in the infinite future (see Fig. \ref{fig5}). Under time reversal the cyclic solution evolves from $\phi=1$ to $\phi=0$.

Things do not change when exploring cross dualities. We can try to see what happens, say, for the GB dual of a RS cosmology ($\theta=1$, $\dth=-1$). Starting from Eq. (\ref{rs}), one gets Eq. (\ref{GBdual}) with $p_0 \to -p_0$, modulo an irrelevant positive constant. The image of the function $\phi(\bar{t})$ is either $\{\phi<1\}$ or $\{\phi>1\}$.

\begin{figure}
\includegraphics[width=8.6cm]{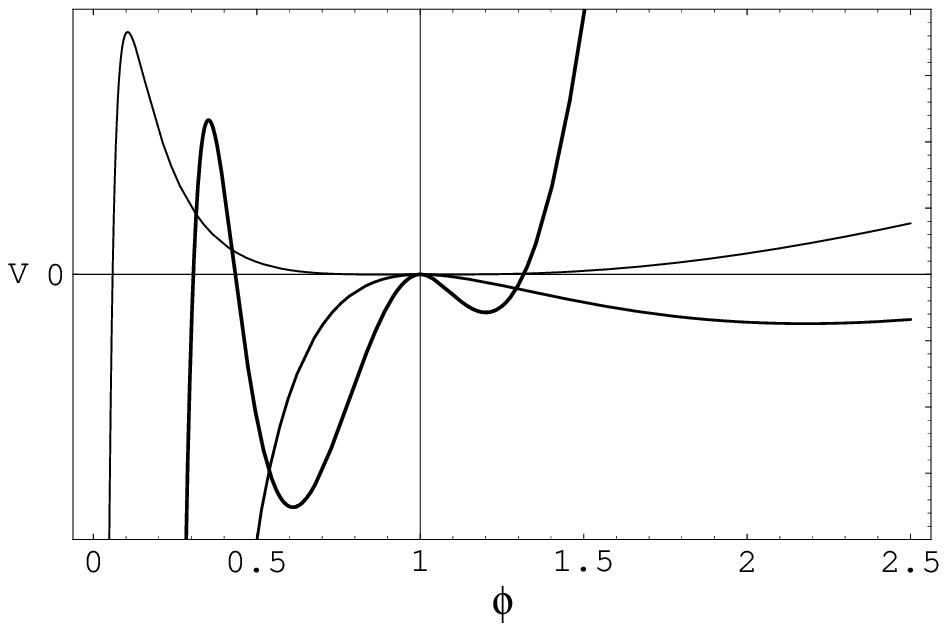}
\caption{\label{fig6}
Gauss-Bonnet potential dual to GB power-law inflation under the mapping (\ref{map2}), for some values of $n$ and $p_0$. The region with $\phi<1$ ($\phi>1$) corresponds to duals with $p_0>0$ ($p_0<0$).}
\end{figure}

Dual potentials can be obtained via the dual of Eq. (\ref{Vphi}) or (\ref{Vtac}). Figure \ref{fig6} shows the potential corresponding to the cosmology Eq. (\ref{GBdual}). Depending on the choice of the parameters $n$ and $p_0$, the function $\bar{V}(\phi)$ has a number of local minima and maxima, can assume negative values, and also be unbounded from below.

The properties of cosmological potentials change very interestingly when going from the 4D picture to the braneworld. Take as examples fast-roll inflation with a standard scalar field and negative potentials \cite{neg}. Fast-roll inflation occurs by definition when the kinetic energy of the scalar field is small with respect to the potential energy, $\dot{\phi}^2 \gg V(\phi)$. In this case one obtains a stiff equation of state ($p=\rho$) and a regime described by
\be
a \sim t^{1/3}\,,\qquad \dot{\phi}^2 \sim t^{-2}\,,\qquad \phi \sim \ln t\,,
\ee
from the Klein-Gordon and Friedmann equations. This implies that at early times the kinetic term dominates over any monomial potential energy $V=\phi^m$. In particular, under these conditions the behaviour of the singularity will depend on the kinetic energy regardless of the choice of the potential. In a generic patch with $\theta \neq 0$, the fast-roll regime is given by
\be
a \sim t^{1/(3q)}\,,\qquad \dot{\phi}^2 \sim t^{\theta-2}\,,\qquad \phi \sim t^{\theta/2}\,.
\ee
Thus the scalar field evolves quite differently in the RS ($\theta=1$) and GB ($\theta=-1$) case. Near the origin, $t\sim 0$, the fast-roll regime is achieved for any $\theta \neq 0$ when $m=2$ and for $\theta>4/(2-m)$ when $m>2$. Therefore the behaviour of the singularity may depend nontrivially on both the contributions of the energy density for suitable (and still simple) potentials on a brane (see also \cite{SST}).

Another result in four dimensions is that potentials with a negative global minimum do not lead to an AdS spacetime. According to the Friedmann equation $H^2=\rho$, the energy density cannot assume negative values; therefore at the minimum $V_\text{min}<0$ the scalar field does not oscillate and stop but increases its kinetic energy until this dominates over the potential contribution. Then one can describe the instability at the minimum in the fast-roll approximation through the only kinetic tern; the Hubble parameter vanishes and becomes negative [so that $(\dot{\phi}^2/2)^\cdot>0$], and the Universe undergoes a bounce.

In a braneworld scenario this might not be the case. In fact, in the RS brane the Friedmann equation is $H^2=\rho [1+\rho/(2\lambda)]$. If the negative minimum is larger than the brane tension, $|V_\text{min}| \gtrsim \lambda$, then, after an eventual fast-roll transition, the quadratic correction dominates near the minimum and $H^2 \approx \rho^2$. The scalar field can relax without spoiling the constraints from the equations of motion.

All that we have said can be investigated in greater detail by means of phase portraits in the three-dimensional space $(\phi,\dot{\phi},H)$. Here we shall not explore the subject further and limit ourselves to the above qualitative comments, whose aim was to stress that complicated dual potentials cannot be discarded by general classical or semiclassical considerations. Rather, from one side they should be studied case by case; from the other side, one or more local features encountered by the scalar field during its evolution could induce interesting phenomena at the quantum level, for instance triggering premature reheating or a series of quantum tunnelings.

\subsubsection{Self-dual solutions and power-law expansion: tachyon case}

In the tachyonic case, from Eq. (\ref{Hdual}) we have $\bar{H}^2=(-2q\ln a)^{-1}$, with $p_0=3/2$ for convenience; in order to have a real Hubble parameter with positive $q$, the dual solution corresponds to the time region with $a<1$. The tachyon solution ($\theta=2$) to Eq. (\ref{self}) is $a(t) = \exp \sqrt[3]{9t^2/(8q)}$.

\begin{figure}[ht]
\includegraphics[width=8.6cm]{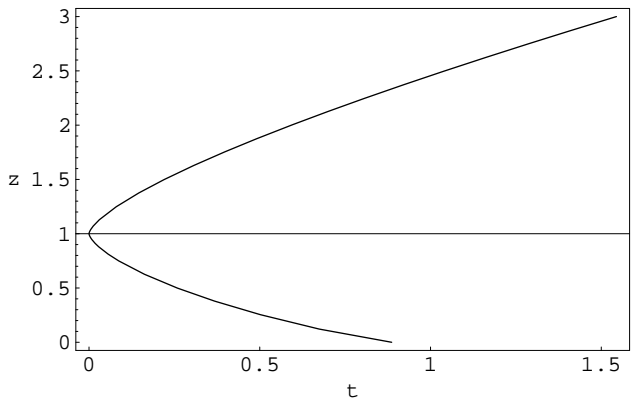}
\caption{\label{fig7}
Numerical plot of the function $z(\bar{t})$ describing the tachyonic cosmologies dual to power-law tachyon inflation. The solid line divides the solutions of the two dualities at $z=1$.}
\end{figure}

\begin{figure}
\includegraphics[width=8.6cm]{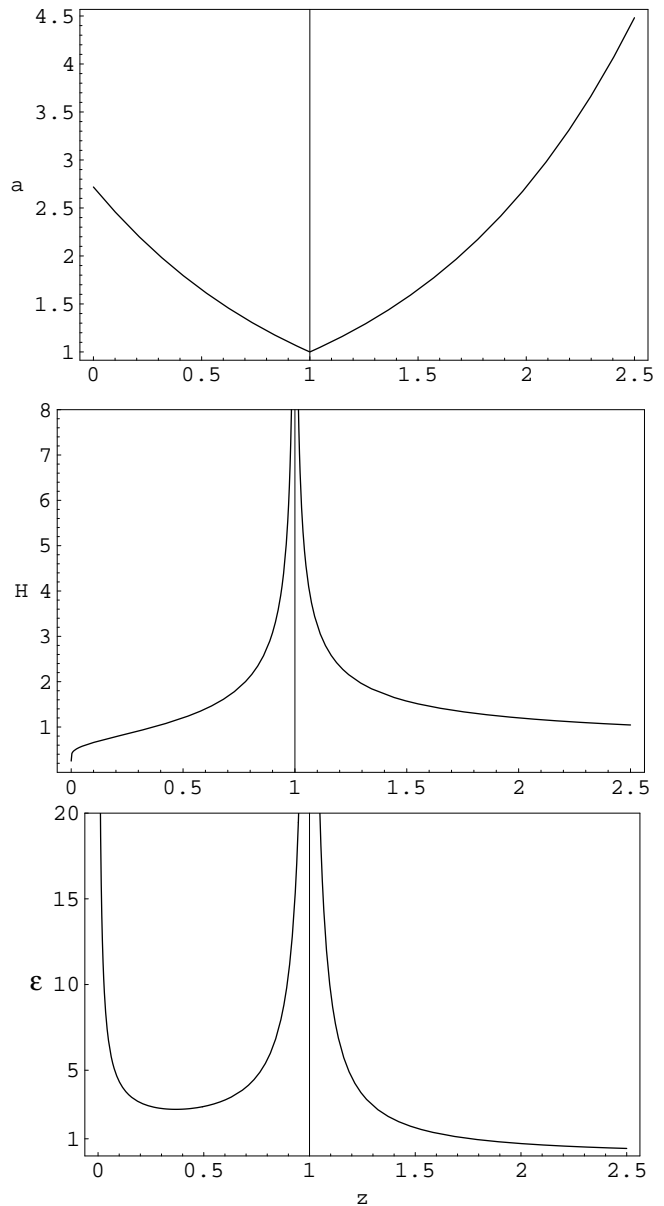}
\caption{\label{fig8}
Tachyon cosmology dual to power-law tachyon inflation for arbitrary positive values of $q$ and $n$. Each panel corresponds to the behaviour of $\bar{a}(z)$, $\bar{H}(z)$, and $\bar{\epsilon}(z)$ (from top to bottom).}
\end{figure}

Power-law inflation is achieved with a tachyon profile
\be
T(t)=T_0 t\,,\qquad V(T)=V_0 T^{\theta-2}\,,
\ee
and
\be
a(T)=(T/T_0)^n\,,\qquad H(T)=nT_0/T\,,
\ee
for all $q$, where $T_0=\sqrt{2/(3qn)}$. Defining $z\equiv (T/T_0)^2$, Eq. (\ref{t}) gives $\dot{z}\propto -(-\ln z)^{-1/2}$, and we get a real dual solution provided $0<z<1$. Since $z$ is a monotonic function of time (see Fig. \ref{fig7} for $z<1$), we express the dual quantities in terms of $z$ itself:
\bs\ba
\bar{a} &=& \exp[-z/(2qn^2)]\,,\\
|\bar{H}| &=& (-\bar{q}n\ln z)^{-1/2}\,,\\
\bar{\epsilon} &=& qn(-z\ln z)^{-1}\,.
\ea\es
Consistently with Eq. (\ref{epsidual}), the dual Hubble radius decreases with time and in fact the dual cosmology decelerates ($\bar{\epsilon}>1$). Figure \ref{fig8} shows the behaviour of the found solution, together with the dual with $p_0=-3/2$.

\subsubsection{Phantom and $q$-duality}

Another duality relates standard solutions to phantom ($\hat{l}=-1$) superinflationary ($\hat{\epsilon}<0$) cosmologies through Eq. (\ref{map2}). For $p_0=-3/2$, one has
\bs\label{mapph}\ba
\hat{a}(\vp)&=&1/y(\vp)\,,\\
\hat{y}(\vp)&=&a(\vp)\,.
\ea\es
Again, there is also a cyclic branch corresponding to a contracting standard phase. The mapping (\ref{mapph}) together with Eq. (\ref{tnorm}) gives $\hat{t}=-\bar{t}$, and we can get the phantom dual solution from the cyclic-dual one:
\bs\ba
\hat{\vp}(t) 			&=& \bar{\vp}(-t)\,,\\
\hat{a}(t) 				&=& \bar{a}(-t)^{-1},\\
\hat{H}(t)        &=& \bar{H}(-t)\,,\\
\hat{\epsilon}(t) &=& -\bar{\epsilon}(-t)\,.
\ea\es
One may realize a similar evolution with superaccelerating scale factor by preserving the null energy condition ($\ell\equiv 1$) and flipping the sign of $q$. The mapping we impose is then
\bs\ba
q_*&=&-q\,,\\
\theta_*&=&4-\theta\,,\\
\epsilon_*&=&-\epsilon\,.
\ea\es
The effect of this correspondence is also clear from Eqs. (\ref{ve}) and (\ref{Hdual}). Actually, the choice of the sign of $q$ determines whether the dual solution is superaccelerating or not. Since the region with $q<0$ generates a phantom cosmology, the name ``cyclic'' often adopted for the transformation (\ref{map2}) with $p_0>0$ is therefore misleading in a braneworld scenario with $q<0$. Same considerations hold for the ``phantom'' mapping, which in this case would generate a solution without phantoms.

Sometimes we will say that cosmologies with $q<0$ mimic scenarios with phantom matter; by this we refer to the above matching of the Hamilton-Jacobi equations and do not mean that there is an effective equivalence between the two, since in the first case the energy density decreases when the scale factor expands, while in the phantom case the energy density increases with $a$.


\subsection{Regular dualities}

It is not yet clear whether the dual solutions constructed so far, especially those with $p_0>0$, describe reasonable (not to mention viable) scenarios. At this point there are two possibilities. The first one is to accept these non-superaccelerating cosmologies and try to explain them by means of some deeper and still missing theoretical ingredient. The second one is to consider their exotic behaviour as a signal that we cannot impose $p_0>0$ (or even $p=$ const) consistently in pure high-energy braneworlds (at least in the RS and GB cases), while the 4D cosmology can be dual to another 4D cosmology. This is due to the fact that the functions $a(\vp)$ and $y(\vp)$ live in different real image sets. Then a new path to follow is to find some mechanism which ``regularizes'' the dual solutions at the asymptotic past and future. The only degree of freedom we could exploit is given by the parameter $p$, which by this line of reasoning must depend on $\vp$. Therefore we are forced to assume Eq. (\ref{p}), which generates the transformation
\bs\label{map3}\ba
\da(\vp)&=&a(\vp)^{p_0}\,,\label{map3a}\\
\dy(\vp)&=&y(\vp)^{1/p_0}\,.
\ea\es
For $\dth\neq0\neq\theta$ the other dual quantities read
\bs \label{altdu1}\ba
\dH(\vp) &=& \left(\frac{p_0 \dalp}{\alpha}\right)^{1/\dth} H(\vp)^{\theta/\dth}\,,\\
\bar{\ve}(\vp) &=& \frac{\ell\bar{\ell}}{p_0^2}\,\ve(\vp) \quad\Rightarrow\\
\de(\vp) &=& \frac{\theta}{\dth p_0}\,\epsilon(\vp)>0\,,
\ea\es
while
\be \label{altdu2}
\bar{t} = p_0 \left(\frac{\alpha}{p_0\dalp}\right)^{1/\dth}\int^\vp d\vp \frac{(\ln a)'}{H^{\theta/\dth}}\,,%
\ee
so that $\bar{\vp}(t)=\vp(t)$ when $\dth=\theta$. 

In 4D ($\dth=\theta=0$), $\dH(\vp)=H(\vp)^{\ell\bar{\ell}/p_0}$, $\bar{t}=p_0\int^\vp d\vp (\ln a)'/H^{\ell\bar{\ell}/p_0}$, and $\de =\ell\bar{\ell}\epsilon/p_0^2$. The cross duality between the general-relativistic framework ($\theta=0$) and a high-energy braneworld ($\dth\neq 0$) is, after a time redefinition,
\bs\ba
\dH(\vp) &=& [\ln H(\vp)]^{-1/\dth}\,,\\
\de(\vp) &=& -\epsilon(\vp)[p_0\dth\ln H(\vp)]^{-1}\,,\\
\bar{t}  &=& p_0 \int^\vp d\vp\, (\ln H)^{1/\dth}(\ln a)'\,.
\ea\es
Clearly, the effect of Eqs. (\ref{map3}), (\ref{altdu1}), and (\ref{altdu2}) results in a rescaling of time when $\dth=\theta$, as one can verify by making the substitution
\be
p_0 \to p_0 \frac{\ln \phi^2}{\phi^2}\,,
\ee
in the RS and GB power-law duals, Eqs. (\ref{RSdual}) and (\ref{GBdual}). In this case (which includes tachyon-tachyon dualities) duals without phantoms are achieved as long as $p_0>0$. 

When $\dth\neq\theta$, this transformation relates the dynamics of different braneworld scenarios. According to the cross duality between RS and GB standard inflation, the dual solution does not superaccelerate if, and only if, $p_0<0$. The power-law case is trivial since the dual GB solution is 
\be
\da = \phi^{-2\bar{n}}\,,\qquad \dH = \phi^2\,,\qquad \de = \bar{n}^{-1}\,,
\ee
where $\bar{n}\equiv -np_0$ and $\phi(t) \propto t^{-1/2}$. 

In the power-law case the mapping (\ref{map3}) can be realized also by
\bs\label{map4}\ba
\da(\vp) &=& [\ln y(\vp)]^s\,,\\
\dy(\vp) &=& \exp\left(\frac{1}{s\theta}\int^\vp d\vp \frac{H}{H'}\right)\,,
\ea\es
where $s$ is a real constant, giving a power-law dual $\da = t^{|s|}$. Note the domain range of the dual scale factor. The dual parameter $\de$ is
\be
\de = \frac{\alpha}{\dalp\dth\theta s^2}\frac{|\dH|^{\dth}}{|H|^\theta}\frac{1}{\epsilon}\,,
\ee
which shows how in general the mapping (\ref{map4}) is not equivalent to Eq. (\ref{map3}). This can be seen also by considering the action of the former in four dimensions, where the dual Hubble parameter reads
\be
\dH = \exp\left[-\frac{3\bar{\ell}}{2 s}\int^\phi d\phi\frac{\ln H}{(\ln H)'}\right]\,.
\ee
The 4D dual of the power-law solution (\ref{pls}) is $\da =\phi^s$, $\dH =\exp (-\phi^2/s)$, and $\de=2\phi^2/s^2$, with potential $\bar{V}=(1-\de/3)\exp(-s\de)$. If $s<0$, there is an instability as $\phi \to \infty$, while for positive $s$ the potential has a local minimum at $\de_*=3+1/s$ [being $V''(\de_*) \propto s$] and vanishes at large $\phi$. 

One can devise other transformations of the Hamilton-Jacobi equation than Eqs. (\ref{map2}), (\ref{map3}), and (\ref{map4}). The last example we give is the following:
\bs\label{map5}\ba
\da(\vp) &=& \exp\left(-\frac{1}{r}\int^\vp \frac{d\vp}{a'}\right)\,,\\
\dy(\vp) &=& \exp [r a(\vp)]\,,
\ea\es
where $r$ is a real constant. For $\dth \neq 0\neq \theta$, the basic equations are
\be
|\dH|=\left(\frac{\dalp}{ra}\right)^{1/\dth},\qquad \de= -\frac{r}{\dth}\frac{a'^2}{a}\,,\qquad \dot{\bar{\vp}}=-ra'\left(\frac{\dalp}{ra}\right)^{1/\dth}.
\ee
The RS$\to$RS dual ($r<0$) has 
\bs\ba
\da &\sim& \exp t^{1-n}\,,\\
\dH &\sim& t^{-n}\,,\\
\de &\sim& t^{n-1}\,.
\ea\es
The RS$\to$GB dual ($r>0$) has 
\bs\ba
\da &\sim& \exp t^{(1-n)/(1-2n)}\,,\\
\dH &\sim& t^{n/(1-2n)}\,,\\
\de &\sim& t^{(1-n)/(1-2n)}\,.
\ea\es
The GB$\to$GB dual ($r>0$) has 
\bs\ba
\da &\sim& \exp t^{(1+n)/(1+2n)}\,,\\
\dH &\sim& t^{-n/(1+2n)}\,,\\
\de &\sim& t^{-(1+n)/(1+2n)}\,. 
\ea\es
In the limit $n \to \infty$, the GB dual of both RS and GB cosmology is $\da \sim \exp \sqrt{t}$, that is the Randall-Sundrum self-dual solution with respect to Eq. (\ref{map2}).

We conclude with an interesting remark. The above dualities connect not only different braneworlds with the same type of scalar field but also patches with different scalars. If one wishes to construct cosmologies with a DBI tachyon, it is sufficient to start from a generic scenario $(\vp,q,\wteta)$ and hit the dual $(T,\bar{q},\dth=2)$ via either Eq. (\ref{map2}), (\ref{map3}), (\ref{map4}), or (\ref{map5}). In particular, with Eq. (\ref{map3}) 
\ba
\dH(T) &=& H(\vp \to T)^{\theta/2}\,,\qquad\qquad \theta\neq 0\,,\\
\dH(T) &=& [\ln H(\vp \to T)]^{-1/2}\,,\qquad \theta= 0\,,
\ea
in agreement, e.g., with previous results on power-law standard and tachyon inflation (see \cite{cal3} and references therein).


\section{Remarks on cosmologies with $q<0$} \label{qbounce}

As discussed in the introduction, when considering a five-dimensional braneworld, bulk moduli modify the Friedmann equation on the brane. In general, to a given orbifolded 5D spacetime and matter source confined on the brane there will correspond a set of junction conditions determining the matter-gravity interaction at the brane position. Conversely, one can always construct a bulk stress tensor such that Eq. (\ref{FRW}) holds for some $q$ \cite{CF}; this is because the junction conditions have enough ($q$-dependent) degrees of freedom at a fixed slice in order to arrange a suitable expansion. In fact, the braneworld alone is not sufficient to fully determine the observable physics and some fundamental principle (e.g., AdS/CFT correspondence) should be advocated from the outside in order to sweep all ambiguities away \cite{NO}. Dealing not with such elegant principles, we shall keep the following discussion on a very qualitative level since the arguments are not yet motivated by a robust theoretical background, if any. 

If the bulk moduli vary with time, the resulting Friedmann evolution on the brane changes accordingly and can be written as in Eq. (\ref{FRW}) but with a time-dependent exponent $q(t)$, at least in a small time interval and under particular energy approximations. The SR parameter $\epsilon$ would not be constant even in the case of constant index of state $w$, see Eq. (\ref{epsi2}). We will make some considerations on the possibilities of nonstandard cosmologies with negative $q$. We stress once again that it is left to see whether such a moduli evolution can be consistently implemented in string theory. A sensible treatment of the moduli sector is crucial for a clear understanding of string cosmology; concrete examples have been constructed, e.g., in \cite{strif}. Nonetheless, a few preliminary remarks might trigger some research in this direction.

To the author's knowledge, so far there has been found only one explicit cosmological model with negative $q$. This comes from higher-derivative gravity theories \cite{ABF1}, in which one can consider a class of gravitational actions like
\be\label{higher}
S_g=\int d^4x \sqrt{g} f(R)\,,
\ee
where $f(R)$ is an arbitrary function of the Ricci scalar and $g$ is the determinant of the metric. It turns out that one can construct suitable expressions for $f(R)$ and get Eq. (\ref{FRW}) in the appropriate limit. The case $f(R)=R -(\sinh R)^{-1}$ is of particular interest, since in the limit of small curvature $R$ (late times, low energy) one gets $H^2 \approx \rho^{-1}$ and can explain the present (super?)acceleration of the Universe. 


\subsection{$q$-bounce?}

In the end of the last section we have seen that there is a formal duality, similar to the ``phantom'' duality, relating standard expanding solutions with $q>0$ to superaccelerating cosmologies with $q<0$. Now it would be interesting to see what are the properties of these solutions and whether they can play some role in bouncing scenarios, as true phantom components may do. For this reason, let us assume that (i) the moduli variation is such that a contracting period with $q<0$ is smoothly followed by a standard $q>0$ expansion and (ii) some stabilization mechanism is effective after the shift in the moduli space, so that the cosmological expansion on the brane can be described by one of the previous models (4D, RS, GB) at sufficiently late times. In the simplest toy model, we can consider a sharp transition from $-q$ to $q$ at the big bang, with $0<q=\text{const}\ll 1$ around the bounce.

\begin{figure}
\includegraphics[width=8.6cm]{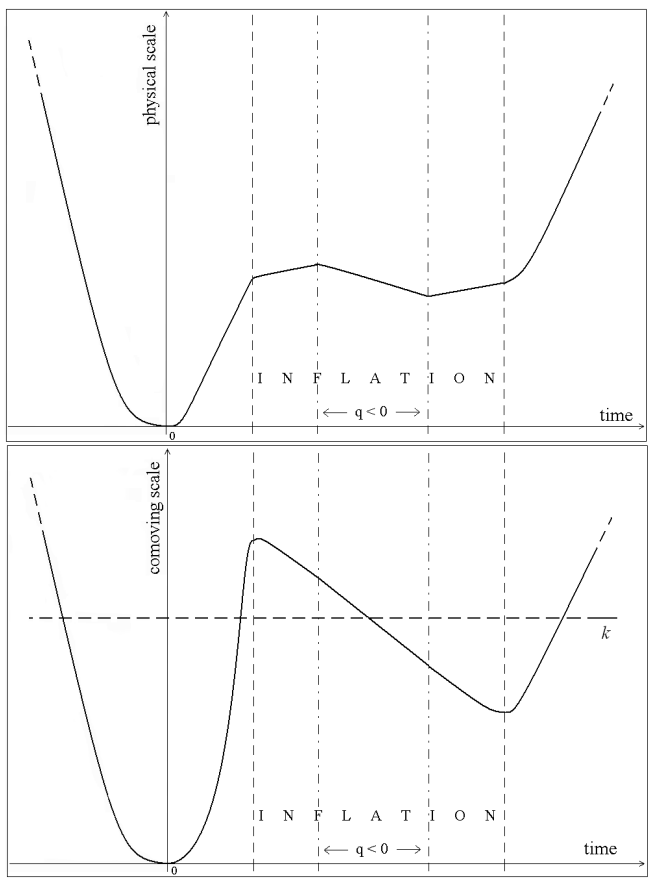}
\caption{\label{fig9}
Inflationary expansion with a short $q\to -q \to q$ transition. The physical Hubble length $|R_H|$ (upper panel) and the comoving one $|R_H|/a$ (lower panel) are plotted in arbitrary units of time. $k$ denotes the comoving wave number of a perturbation exiting the horizon during the $q$-transition. For $t<0$ a standard contracting behaviour is represented.}
\end{figure}

The contracting phase is deflationary since $\ddot{a}>0$ and is actually superaccelerating if the brane content is not phantomlike. The absolute value of the Hubble rate decreases to zero while the energy density $\rho(t)$ approaches the singularity at $\rho(0)=\infty$; in a standard contracting phase (Fig. \ref{fig9} for $t<0$) it is the Hubble radius that decreases.

Note that at neither this nor any other stage we are saying anything about ``the creation of the Universe,'' since all these considerations regard the cosmological evolution from a brane-observer point of view rather than the global spacetime structure. Although the backreaction on the brane is governed by the moduli evolution, the braneworld as a geometrical object does not undergo any dramatical transition and is considered to be present at any time in order to make sense of the modified Friedmann equation before, during, and after the bounce. Genuine braneworld creation has been considered recently in \cite{AM}.

There are several advantages in constructing a model of bounce with varying $q$. First, it avoids the reversal problem due to the monotonicity of the Hubble parameter in general relativity \cite{KOSST}. Second, one does not encounter the classical instabilities of background contracting solutions with $w=\text{const}$ found in \cite{GKST,EWST}. In this case, from the continuity equation the energy density scales as $\rho =a^{-3(1+w)}$, up to some constant factor. In a contracting universe, if $w \lesssim -1$ the energy density of the scalar field is nonincreasing, while an extra matter or radiation component increases with time. Therefore solutions with $w \lesssim -1$ are not attractors as regards the isotropic cosmological evolution, while solutions with $w > 1$ ($\gg 1$ in cyclic or ekpyrotic scenarios) are stable. 

Put into another way, for constant $w$ one has $R_H \sim t$ and $a \sim t^{1/\epsilon}$. When $0<\epsilon<1$, $a$ grows more rapidly than the Hubble radius and quantum fluctuations can leave the horizon; for $q>0$, $\epsilon>1$, and $H<0$, a necessary condition for getting a scale-invariant spectrum is that $R_H$ shrinks more rapidly than $a$, that is $\epsilon >1$. When $q<0$, the scale factor shrinks as the Hubble radius decreases and vice versa, and no apparent critical index of state is required.

Since there is no concrete model motivating a patch transition, this scenario is not less arbitrary than those invoking an \emph{ad hoc} phantom matter. Matter with $w<-1$ has been advocated both in the context of bouncing cosmologies and for explaining modern data on cosmic acceleration. Although it has been criticized in many respects \cite{crit} and is not strictly necessary to bring current observations to account \cite{HN}, a phantom component still can be embedded in string theory (e.g., \cite{fra02}) and has attractive features; for instance, in a cyclic phantom universe black holes are tore apart and are prevented to cannibalize the cosmological horizon during one of the contracting phases \cite{BFK,BDE}. Of course this is not the case for $q$-cosmologies in which the null energy condition, determining the evolution of the black hole mass, is preserved.

Another clear shortcoming is that there is no apparent reason why the scale factor should reverse its evolution exactly during the $q$-bounce. Therefore there is no immediate relation between solutions with negative $q$ and bouncing models of the early Universe. Anyway Eq. (\ref{FRW}) is only a particular case of a wider and more realistic class of cosmological evolutions, to which the RS scenario itself does belong. If the nonstandard behaviour of the 4D Friedmann equation arises as a correction to the linear term, then it is natural to write it down as a polynomial (rather than a monomial) in $\rho$:
\be
H^2=b_1\rho^{q_1}-b_2\rho^{q_2}\,,
\ee
where $q_1,q_2,b_1,b_2$ are constants; one can always set one of the $b_i$'s to 1 in appropriate units. In the RS two-brane case $q_1=1$, $b_1=1$, and $q_2=2$, while $b_2=-(2\lambda)^{-1}$ in the type 2 model (matter on the brane with positive tension) and $b_2=(2|\lambda|)^{-1}$ in the type 1 model (matter on the brane with negative tension). If $b_1,b_2>0$, then a bounce occurs at
\be \label{rhob}
\rho_b\equiv(b_2/b_1)^{1/(q_1-q_2)}\,.
\ee
Under the additional assumption that $\text{sgn}(q_1)\neq \text{sgn}(q_2)$, a period of nonphantom superacceleration may dominate at some point of the evolution, according to the sign of the coefficients. But about this we will say no more.


\subsection{$q$-bump?}

Another possibility arises when the evolution of the moduli in the bulk is such that $q$ changes from positive to negative to again positive values in some interval $\Delta t=t_e-t_i$. In the case the transition $q \rightarrow -q \rightarrow q$ happens during the inflationary period, some interesting features in the power spectrum may be generated. A bump in the power spectrum would occur for those perturbations crossing the horizon during the patch transition. In the toy model, $q(t)$ is a step function with sharp transitions and the characteristic time of the event is small with respect to the total duration $\Delta t_\text{inf}$ of the accelerated expansion, $\Delta t/\Delta t_\text{inf}\ll 1$; in Fig. \ref{fig9} the interval $\Delta t$ is exaggerated. Perturbations leaving the horizon during this period will break scale invariance in a comoving wave number interval $\Delta k=k(t_e)-k(t_i)$. 

To get some idea of the properties of the arising feature it is more convenient to consider a smoothly varying $q$, for example with a Gaussian profile centered at some time $t_0$,
\be 
q(t)=q_0-q_1 e^{-(t-t_0)^2/\sigma^2},
\ee
where $q_1>q_0>0$ and $\Delta t \sim \sigma$ is the region of validity of the approximation.

Let us recall that the expressions for the squared scalar and tensor amplitudes and their ratio are ($\beta_q=1$)
\bs\label{qspec}\ba
A_s^2(\vp) &\propto& qH^{2+\theta}/\epsilon\,,\\
A_t^2      &\propto& |q|H^{2+\theta}/\zeta_q\,,\label{At}\\
r          &\equiv&  A_t^2/A_s^2=|\epsilon|/\zeta_q\,,
\ea\es
where $\zeta_q$ is a $O(1)$ coefficient depending on the concrete gravity model and it has been assumed to be positive without loss of generality. Equations (\ref{qspec}) are valid to lowest SR order; in fact, around $q\sim 0$ the parameter $\epsilon \sim 0$ and the SR approximation still holds. The scalar and tensor indices are, near $k_0=k(t_0)$,
\ba
n_s-1 &\equiv& \frac{d \ln A_s^2}{d \ln k} \approx -(2+\theta)\epsilon+(1-\epsilon)\gamma\,,\\
n_t   &\equiv& \frac{d \ln A_t^2}{d \ln k} \approx -(2+\theta)\epsilon\,.
\ea
A negative $q \approx q_0-q_1$ corresponds to $\theta > 2$ and a very blue-tilted gravitational wave spectrum, an effect that has been found in ekpyrotic models also \cite{dual}. However, models with $\gamma > 4\epsilon$ have even a blue-tilted scalar spectrum; if $k_0 \ll 10$, that is at long wavelengths, this might fit with the loss of power in the comic microwave background (CMB) quadrupole region found in recent data. Note that the divergence $\theta \to \infty$ at the bounce is typical of purely adiabatic perturbations. In general relativity, the consistent introduction of entropy perturbations, generated by the mixing modes of a multicomponent fluid, compensates the curvature divergence \cite{PPG} and a similar mechanism might operate in this case, too.


\subsection{$q$-inflation?}

Because of its features one might think to regard an expanding $q<0$ era as a substitute of standard inflation. For example, we can devise a superaccelerating universe filled by a not-slow-rolling scalar field with a generic potential. The expansion inflates the fluctuations of the field (thus explaining the large-scale anisotropies) until the moduli evolution changes the sign of $q$ and gracefully exits to a normal, decelerating expansion. A few properties of expanding $q$-models were already outlined in \cite{cal3}.

One of the most important strongholds of inflation is its capability to select a de Sitter vacuum from a non fine-tuned set of initial conditions. This property is encoded in the definition of the ``inflationary attractor''. If there exists an attractor behaviour such that cosmological solutions with different initial conditions (i.c.) rapidly converge, then the (post-) inflationary physics will generate observables which are independent of such conditions. Let $H_o(\vp)>0$ be a background expanding solution (denoted with the subscript $o$) of the Hamilton-Jacobi equations and consider a linear perturbation $\delta H(\vp)$ which does not reverse the sign of $\dot{\vp}>0$. From the linearized equation of motion for the inflaton field it turns out that \cite{cal3}
\be
\delta H(\vp) = \delta H(\vp_o)\,\exp \int_{\vp_o}^\vp \frac{d\vp}{c_q} \left[(2-\theta)\left(1+\frac{\theta}{3}\,\epsilon\right)\right]\frac{H_o^{\wteta+1}}{H_o'}\,,\nonumber\\\label{attrac}
\ee
where $\beta_q=1$ and $c_q \equiv (2/3q)^2$. All linear perturbations are exponentially damped when the integrand is negative definite and, since $H_o'$ and $\dot{\vp}$ have opposing signs when $q$ is positive, this occurs when the term inside square brackets is positive. When $\theta>2$ ($q<0$), $H_o'$ and $\dot{\vp}$ have concording signs. In this case, linear perturbations are suppressed when $|\epsilon|<3/\theta$; in the large $\theta$ limit, that is when $q$ is close to vanish in the realistic case of smoothly varying moduli, this condition leads to a trivial de Sitter expansion $H=\beta_0$ insensitive of the matter content. Solutions with a greater SR parameter would depend on the initial conditions in an unpleasant way. 

The condition $|\epsilon|<3/2$, though more stringent than those of standard inflationary scenarios with positive $\epsilon$ (4D and RS: $\forall \epsilon$; GB: $\epsilon < 3$), does not severely constrain the dynamics of the scalar field in order to have a sufficiently flat potential, provided not a too negative $q$. However, it is important to stress that this new picture might not replace inflation because of this possible fine tuning, $|q|\ll 1$. Therefore it is not clear whether the dependence on i.c. would survive or not after the bump, although a sufficient amount of $q$-inflation might have erased any memory of the i.c. at this time.


\section{Conclusions} \label{concl}

In this paper we have generalized the four-dimensional triality between inflationary, cyclic, and phantom cosmologies to the case of a braneworld scenario with a modified effective Friedmann evolution $H^2=\rho^q$. The exact, simple relations Eqs. (\ref{4Dduality}) between the parameters $\epsilon=\dot{R}_H$ of models with an ordinary scalar field are broken and extended consequently. The self-dual solutions and the duals of power-law inflation have been provided for three scenarios (General Relativity, Randall-Sundrum, and Gauss-Bonnet braneworlds) in the presence of either a normal scalar field or a Born-Infeld tachyon. Finally, starting from a new version of the ``phantom'' duality, we have set some remarks on cosmologies with $q<0$.

The structure of the triality is deeply modified: The cosmologies dual to inflation either display singularities within finite time intervals or are nonsingular at the origin. This last feature is appealing as regards the construction of nonsingular bounces and would deserve further attention. Hopefully, the cross dualities we have found will help in the understanding of high-energy phantom and bouncing models of the early Universe.

Many topics remain to explore. For example, one should consider the contribution of the nonlocal physics of the bulk in order to set a truly consistent picture of braneworld cosmologies and dualities. In the typical inflationary context, Eq. (\ref{FRW}) encodes the most part of the braneworld effective evolution; in fact, the simplest contribution of the projected Weyl tensor is $\propto a^{-4}$ and is damped away during the accelerated expansion. However, the dark radiation term is no longer negligible in a contracting universe and should be taken into account.

In parallel, it would be interesting to explore two other directions. The first, most important issue should be to motivate $q<0$ scenarios within string theory, super or quantum gravity, since at this stage they are rather speculative. The second direction goes towards a study of the cosmological perturbations through the bounce, by further modeling the too simple step-function transitions we presented. A more concrete model would try to provide a smooth big crunch/big bang phase and allow a nonsharp transition in $q$.
 
In order to fully resolve the singular bounce we should rely on a description more general than classical gravity. To find reasonable solutions of the big bang singularity and embed a bouncing picture in a well-established (string-) theoretical framework will perhaps be one of the most promising lines of research in the following years, not only for the immediate cosmological implications (observability of preinflationary physics and comprehension of the high-energy early Universe) but also because it might lead to a better understanding of the still controversial but intriguing \emph{landscape} of vacua \cite{ban04}.


\begin{acknowledgments}
The author thanks the organizers of Cargèse Summer School 2004, where part of this work was done, for their kind hospitality. I am grateful to Arianna Trevisiol for her patience and support.

\textit{Note added:} After the completion of the first version of this work \cite{calv1} another paper treating similar topics appeared on the electronic preprint server \cite{CLLM}. In particular, their results are in agreement with ours when the duality transformations act on scaling solutions of a single patch with $L=H^{\theta/2}$.
\end{acknowledgments}



\begin{thebibliography}{50}

\bibitem{strin} R. Brandenberger and C. Vafa, Nucl. Phys. \textbf{B316}, 391 (1989).

								A.A. Tseytlin and C. Vafa, Nucl. Phys. \textbf{B372}, 443 (1992) [\eprint{hep-th/9109048}].
								
								N. Kaloper, R. Madden, and K.A. Olive, Nucl. Phys. \textbf{B452}, 677 (1995) [\eprint{hep-th/9506027}].
								
								R. Easther, K. Maeda, and D. Wands, Phys. Rev. D \textbf{53}, 4247 (1996) [\eprint{hep-th/9509074}].

  						  J.E. Lidsey, Phys. Rev. D \textbf{55}, 3303 (1997) [\eprint{gr-qc/9605017}].
  						  
  						  A. Lukas, B.A. Ovrut, and D. Waldram, Phys. Lett. B \textbf{393}, 65 (1997) [\eprint{hep-th/9608195}].
                
								J.C. Fabris, R.G. Furtado, P. Peter, and N. Pinto-Neto, Phys. Rev. D \textbf{67}, 124003 (2003) [\eprint{hep-th/0212312}].
								
								G.T. Horowitz and J. Polchinski, Phys. Rev. D \textbf{66}, 103512 (2002) [\eprint{hep-th/0206228}].
								
								M. Fabinger and J. McGreevy, J. High Energy Phys. 06 (2003) 042 [\eprint{hep-th/0206196}].
								
								S. Tsujikawa, Class. Quantum Grav. \textbf{20}, 1991 (2003) [\eprint{hep-th/0302181}].
								
								T. Hertog and G.T. Horowitz, J. High Energy Phys. 07 (2004) 073 [\eprint{hep-th/0406134}].
								
								N. Turok, M. Perry, and P.J. Steinhardt, Phys. Rev. D \textbf{70}, 106004 (2004) [\eprint{hep-th/0408083}].
								
								G.J. Russo and P.K. Townsend, \eprint{hep-th/0408220}.
								
								T. Hertog, \eprint{hep-th/0409160}.
								
\bibitem{old}   R.C. Tolman, Phys. Rev. \textbf{37}, 1639 (1931).

								R.C. Tolman, Phys. Rev. \textbf{38}, 1758 (1931).

								G. Lema\^{\i}tre, Ann. Soc. Sci. Bruxelles I A \textbf{53}, 51 (1933).

								R.C. Tolman, \textit{Relativity, Thermodynamics and Cosmology} (Clarendon, Oxford, 1934).

								M.J. Rees, Observatory \textbf{89}, 193 (1969).

								R.H. Dicke and P.J.E. Peebles, in \textit{General Relativity: An Einstein Centenary Survey}, edited by S.W. Hawking and W. Israel (Cambridge University Press, Cambridge, England, 1979), p.504.

								Ya.B. Zel'dovich and I.D. Novikov, \textit{Relativistic Astrophysics, Vol. 2, The Structure and Evolution of the Universe} (University of Chicago Press, Chicago, 1983).								
								
\bibitem{BB}    T.J. Battefeld and R. Brandenberger, Phys. Rev. D \textbf{70}, 121302(R) (2004) [\eprint{hep-th/0406180}].  

\bibitem{PP1}	  P. Peter and N. Pinto-Neto, Phys. Rev. D \textbf{65}, 023513 (2002) [\eprint{gr-qc/0109038}].  

\bibitem{CDC}   C. Cartier, R. Durrer, and E.J. Copeland, Phys. Rev. D \textbf{67}, 103517 (2003) [\eprint{hep-th/0301198}].

\bibitem{PPG}   P. Peter, N. Pinto-Neto, and D.A. Gonzalez, J. Cosmol. Astropart. Phys. 12 (2003) 003 [\eprint{hep-th/0306005}].

\bibitem{KST}   J. Khoury, P.J. Steinhardt, and N. Turok, Phys. Rev. Lett. \textbf{92}, 031302 (2004) [\eprint{hep-th/0307132}].

\bibitem{EWST}  J.K. Erickson, D.H. Wesley, P.J. Steinhardt, and N. Turok, Phys. Rev. D \textbf{69}, 063514 (2004)  [\eprint{hep-th/0312009}].

\bibitem{AW}    L.E. Allen and D. Wands, Phys. Rev. D \textbf{70}, 063515 (2004) [\eprint{astro-ph/0404441}].  
                
\bibitem{close} N. Kanekar, V. Sahni, and Y. Shtanov, Phys. Rev. D \textbf{63}, 083520 (2001) [\eprint{astro-ph/0101448}].

								J. Hwang and H. Noh, Phys. Rev. D \textbf{65}, 124010 (2002) [\eprint{astro-ph/0112079}].
								
                C. Gordon and N. Turok, Phys. Rev. D \textbf{67}, 123508 (2003) [\eprint{hep-th/0206138}].
								
								J. Martin and P. Peter, Phys. Rev. D \textbf{68}, 103517 (2003) [\eprint{hep-th/0307077}].
								
								J. Martin and P. Peter, Phys. Rev. Lett. \textbf{92}, 061301 (2004) [\eprint{astro-ph/0312488}].
								
								B.K. Tippett and K. Lake, \eprint{gr-qc/0409088}.
								
\bibitem{BM}    R. Brustein and R. Madden, Phys. Lett. B \textbf{410}, 110 (1997) [\eprint{hep-th/9702043}].

                R. Brustein and R. Madden, Phys. Rev. D \textbf{57}, 712 (1998) [\eprint{hep-th/9708046}].

\bibitem{PP2}	  P. Peter and N. Pinto-Neto, Phys. Rev. D \textbf{66}, 063509 (2002) [\eprint{hep-th/0203013}].

\bibitem{pha1}  R.R. Caldwell, Phys. Lett. B \textbf{545}, 23 (2002) [\eprint{astro-ph/9908168}]. 

\bibitem{cal3}  G. Calcagni, Phys. Rev. D \textbf{69}, 103508 (2004) [\eprint{hep-ph/0402126}].
                
\bibitem{ACL}   J.M. Aguirregabiria, L.P. Chimento, and R. Lazkoz, Phys. Rev. D \textbf{70}, 023509 (2004) [\eprint{astro-ph/0403157}].
								
\bibitem{phant}	Y.-S. Piao and E Zhou, Phys. Rev. D \textbf{68}, 083515 (2003) [\eprint{hep-th/0308080}].

								P.F. Gonz\'{a}lez-D\'{\i}az, Phys. Lett. B \textbf{586}, 1 (2004) [\eprint{astro-ph/0312579}].
								
								Y.-S. Piao and Y.-Z. Zhang, Phys. Rev. D \textbf{70}, 063513 (2004) [\eprint{astro-ph/0401231}].

								O. Bertolami, A.A. Sen, S. Sen, and P.T. Silva, Mon. Not. R. Astron. Soc. \textbf{353}, 329 (2004) [\eprint{astro-ph/0402387}].
								
								J.D. Barrow, Class. Quantum Grav. \textbf{21}, L79 (2004) [\eprint{gr-qc/0403084}].
								
								J. Hao and X. Li, \eprint{astro-ph/0404154}.
								
								Z.-K. Guo, Y.-S. Piao, and Y.-Z. Zhang, Phys. Lett. B \textbf{594}, 247 (2004) [\eprint{astro-ph/0404225}].
								
								M. Bouhmadi-L\'{o}pez and J.A. Jim\'{e}nez-Madrid, \eprint{astro-ph/0404540}.
								
								E. Elizalde, S. Nojiri, and S.D. Odintsov, Phys. Rev. D \textbf{70}, 043539 (2004) [\eprint{hep-th/0405034}].
								
								S. Nojiri and S.D. Odintsov, Phys. Lett. B \textbf{595}, 1 (2004) [\eprint{hep-th/0405078}].
								
								L.P. Chimento and R. Lazkoz, Mod. Phys. Lett. A \textbf{19}, 2479 (2004) [\eprint{gr-qc/0405020}].
								
								A. Gruzinov, \eprint{astro-ph/0405096}.
								
								P.H. Frampton and T. Takahashi, Astropart. Phys. \textbf{22}, 307 (2004) [\eprint{astro-ph/0405333}].
								
								L.P. Chimento and R. Lazkoz, \eprint{astro-ph/0405518}.
								
								S.D.H. Hsu, A. Jenkins, and M.B. Wise, Phys. Lett. B \textbf{597}, 270 (2004) [\eprint{astro-ph/0406043}].
								
								P.F. Gonz\'{a}lez-D\'{\i}az and J.A. Jiménez-Madrid, Phys. Lett. B \textbf{596}, 16 (2004) [\eprint{hep-th/0406261}].
								
								S. Nojiri, \eprint{hep-th/0407099}.
								
								S.K. Srivastava, \eprint{astro-ph/0407048}.
								
								A. Vikman, \eprint{astro-ph/0407107}.

							  E. Babichev, V. Dokuchaev, and Yu. Eroshenko, \eprint{astro-ph/0407190}. 
							  
							  P.H. Frampton, \eprint{astro-ph/0407353}.
							  
							  P.F. Gonz\'{a}lez-D\'{\i}az and C.L. Sig\"{u}enza, Nucl. Phys. \textbf{B697}, 363 (2004) [\eprint{astro-ph/0407421}].
							  
							  P. Wu and H. Yu, \eprint{astro-ph/0407424}.
							  
							  B. Feng, M. Li, Y.-S. Piao, and X. Zhang, \eprint{astro-ph/0407432}.
							  
							  F.C. Carvalho and A. Saa, Phys. Rev. D \textbf{70}, 087302 (2004) [\eprint{astro-ph/0408013}].
							  
							  P.F. Gonz\'{a}lez-D\'{\i}az, \eprint{hep-th/0408225}.
							  
							  W. Fang, H.Q. Lu, Z.G. Huang, and K.F. Zhang, \eprint{hep-th/0409080}.

\bibitem{ekpy1} J. Khoury, B.A. Ovrut, P.J. Steinhardt, and N. Turok, Phys. Rev. D \textbf{64}, 123522 (2001) [\eprint{hep-th/0103239}].

								R. Kallosh, L. Kofman, and A. Linde, Phys. Rev. D \textbf{64}, 123523 (2001) [\eprint{hep-th/0104073}].

								R. Kallosh, L. Kofman, A. Linde, and A. Tseytlin, Phys. Rev. D \textbf{64}, 123524 (2001) [\eprint{hep-th/0106241}].

\bibitem{KOSST} J. Khoury, B.A. Ovrut, N. Seiberg, P.J. Steinhardt, and N. Turok, Phys. Rev. D \textbf{65}, 086007 (2002) [\eprint{hep-th/0108187}].

\bibitem{ekpy2} R. Brandenberger and F. Finelli, J. High Energy Phys. 11 (2001) 056 [\eprint{hep-th/0109004}].

								D.H. Lyth, Phys. Lett. B \textbf{526}, 173 (2002) [\eprint{hep-ph/0110007}].
								
								S. Tsujikawa, Phys. Lett. B \textbf{526}, 179 (2002) [\eprint{gr-qc/0110124}].

								P.J. Steinhardt and N. Turok, Phys. Rev. D \textbf{65}, 126003 (2002) [\eprint{hep-th/0111098}]. 
								
								J. Martin, P. Peter, N. Pinto-Neto, and D.J. Schwarz, Phys. Rev. D \textbf{65}, 123513 (2002) [\eprint{hep-th/0112128}].

                J. Khoury, B.A. Ovrut, P.J. Steinhardt, and N. Turok, Phys. Rev. D \textbf{66}, 046005 (2002) [\eprint{hep-th/0109050}].
                
                R. Durrer and F. Vernizzi, Phys. Rev. D \textbf{66}, 083503 (2002) [\eprint{hep-ph/0203275}].

								S. Tsujikawa, R. Brandenberger, and F. Finelli, Phys. Rev. D \textbf{66}, 083513 (2002), [\eprint{hep-th/0207228}].
								
								J. Martin, P. Peter, N. Pinto-Neto, and D.J. Schwarz, Phys. Rev. D \textbf{67}, 028301 (2003) [\eprint{hep-th/0204222}].

                A.J. Tolley, N. Turok, and P.J. Steinhardt, Phys. Rev. D \textbf{69}, 106005 (2004)
 [\eprint{hep-th/0306109}].
                 
                P.J. Steinhardt and N. Turok, \eprint{astro-ph/0404480}.	
																          											  
\bibitem{GKST}  S. Gratton, J. Khoury, P.J. Steinhardt, and N. Turok, Phys. Rev. D \textbf{69}, 103505 (2004) [\eprint{astro-ph/0301395}].

\bibitem{dual}  J. Khoury, P.J. Steinhardt, and N. Turok, Phys. Rev. Lett. \textbf{91}, 161301 (2003) [astro-ph/0302012].
                
                L.A. Boyle, P.J. Steinhardt, and N. Turok, Phys. Rev. D \textbf{70}, 023504 (2004) [\eprint{hep-th/0403026}].
                
                Y.-S. Piao and Y.-Z. Zhang, Phys. Rev. D \textbf{70}, 043516 (2004) [\eprint{astro-ph/0403671}].
                
                Y.-S. Piao, \eprint{hep-th/0404002}.
                
\bibitem{lid04} J.E. Lidsey, Phys. Rev. D \textbf{70}, 041302 (2004) [\eprint{gr-qc/0405055}].

\bibitem{phdu}  L.P. Chimento, Phys. Rev. D \textbf{65}, 063517 (2002).
              
                J.M. Aguirregabiria, L.P. Chimento, A.S. Jakubi, and R. Lazkoz, Phys. Rev. D \textbf{67}, 083518 (2003) [\eprint{gr-qc/0303010}].
                
                L.P. Chimento and R. Lazkoz, Phys. Rev. Lett. \textbf{91}, 211301 (2003) [\eprint{gr-qc/0307111}].
                
							  M.P. D\c{a}browski, T. Stachowiak, and M. Szyd{\l}owski, Phys. Rev. D \textbf{68}, 103519 (2003) [\eprint{hep-th/0307128}].

\bibitem{altdu} R. Brustein, M. Gasperini, and G. Veneziano, Phys. Lett. B \textbf{431}, 277 (1998) [\eprint{hep-th/9803018}].

                D. Wands, Phys. Rev. D \textbf{60}, 023507 (1999) [\eprint{gr-qc/9809062}].    
							  									
\bibitem{BKM}   J.D. Barrow, D. Kimberly, and J. Magueijo, Class. Quantum Grav. \textbf{21}, 4289 (2004) [\eprint{astro-ph/0406369}].

\bibitem{MMMZ}  M. Maceda, J. Madore, P. Manousselis, and G. Zoupanos, Eur. Phys. J. C \textbf{36}, 529 (2004) [\eprint{hep-th/0306136}].		
					      
\bibitem{quang}	P. Singh and A. Toporensky, Phys. Rev. D \textbf{69}, 104008 (2004) [\eprint{gr-qc/0312110}].

							  J.E. Lidsey, D.J. Mulryne, N.J. Nunes, and R. Tavakol, Phys. Rev. D \textbf{70}, 063521 (2004) [\eprint{gr-qc/0406042}].

								M. Bojowald, R. Maartens, and P. Singh, Phys. Rev. D \textbf{70}, 083517 (2004) [\eprint{hep-th/0407115}].
								
								G. Date and G.M. Hossain, \eprint{gr-qc/0407074}.
								
\bibitem{sri04}	S.K. Srivastava, \eprint{hep-th/0404170}.							
								
\bibitem{SS}    Y. Shtanov and V. Sahni, Phys. Lett. B \textbf{557}, 1 (2003) [\eprint{gr-qc/0208047}].

\bibitem{BFK}   M.G. Brown, K. Freese, and W.H. Kinney, \eprint{astro-ph/0405353}.

\bibitem{PZ}    Y.-S. Piao and Y.-Z. Zhang, \eprint{gr-qc/0407027}.
                 
\bibitem{GW}		W.D. Goldberger and M.B. Wise, Phys. Rev. Lett. \textbf{83}, 4922 (1999) [\eprint{hep-ph/9907447}].

								W.D. Goldberger and M.B. Wise, Phys. Lett. B \textbf{475}, 275 (2000) [\eprint{hep-ph/9911457}].
								
\bibitem{quco}  C. Cs\'{a}ki, M. Graesser, C. Kolda, and J. Terning, Phys. Lett. B \textbf{462}, 34 (1999) [\eprint{hep-ph/9906513}].

								J.M. Cline, C. Grojean, and G. Servant, Phys. Rev. Lett. \textbf{83}, 4245 (1999) [\eprint{hep-ph/9906523}].         
                
						    P. Bin\'{e}truy, C. Deffayet, and D. Langlois, Nucl. Phys. \textbf{B565}, 269 (2000) [\eprint{hep-th/9905012}].
						    
						    P. Bin\'{e}truy, C. Deffayet, U. Ellwanger, and D. Langlois, Phys. Lett. B \textbf{477}, 285 (2000) [\eprint{hep-th/9910219}].
						    
								\'{E}.\'{E}. Flanagan, S.-H.H. Tye, and I. Wasserman, Phys. Rev. D \textbf{62}, 044039 (2000) [\eprint{hep-ph/9910498}].										            

\bibitem{gabo}   J.E. Kim, B. Kyae, and H.M. Lee, Phys. Rev. D \textbf{62}, 045013 (2000) [\eprint{hep-ph/9912344}].

								 J.E. Kim, B. Kyae, and H.M. Lee, Nucl. Phys. \textbf{B582}, 296 (2000); \textbf{B591}, 587(E) (2000) [\eprint{hep-th/0004005}].
								 
						     S. Nojiri and S.D. Odintsov, J. High Energy Phys. \textbf{07}, 049 (2000) [\eprint{hep-th/0006232}].
						     							   							  				   
					       C. Charmousis and J. Dufaux, Class. Quantum Grav. \textbf{19}, 4671 (2002) [\eprint{hep-th/0202107}].
					          				   
  							 S. Nojiri, S.D. Odintsov, and S. Ogushi, Phys. Rev. D \textbf{65}, 023521 (2002) [\eprint{hep-th/0108172}].
  							 
						     J.E. Lidsey, S. Nojiri, and S.D. Odintsov, J. High Energy Phys. 06 (2002) 026 [\eprint{hep-th/0202198}].
						     
							   S. Nojiri, S.D. Odintsov, and S. Ogushi, Int. J. Mod. Phys. A \textbf{17}, 4809 (2002) [\eprint{hep-th/0205187}].
							   
						     J.E. Lidsey and N.J. Nunes, Phys. Rev. D \textbf{67}, 103510 (2003) [\eprint{astro-ph/0303168}].
						     
						     J.P. Gregory and A. Padilla, Class. Quantum Grav. \textbf{20}, 4221 (2003) [\eprint{hep-th/0304250}].
						     						   
	 					     E. Gravanis and S. Willison, Phys. Lett. B \textbf{562}, 118 (2003) [\eprint{hep-th/0209076}].
	 					     
	 					     J.F. Dufaux, J.E. Lidsey, R. Maartens, and M. Sami, Phys. Rev. D \textbf{70}, 083525 (2004) [\eprint{hep-th/0404161}].
	 					     
	 					     S. Tsujikawa, M. Sami, and R. Maartens, Phys. Rev. D \textbf{70}, 063525 (2004) [\eprint{astro-ph/0406078}].
						
\bibitem{MFB}   V.F. Mukhanov, H.A. Feldman, and R.H. Brandenberger, Phys. Rep. \textbf{215}, 203 (1992).

\bibitem{CLLM}  E.J. Copeland, S.J. Lee, J.E. Lidsey, and S. Mizuno, \eprint{astro-ph/0410110}.

\bibitem{neg}   A.D. Linde, J. High Energy Phys. 11 (2001) 052 [\eprint{hep-th/0110195}].

                G. Felder, A. Frolov, L. Kofman, and A. Linde, Phys. Rev. D \textbf{66}, 023507 (2002) [\eprint{hep-th/0202017}].

\bibitem{SST}   M. Sami, N. Savchenko, and A. Toporensky, \eprint{hep-th/0408140}.

\bibitem{CF}    D.J.H. Chung and K. Freese, Phys. Rev. D \textbf{61}, 023511 (2000) [\eprint{hep-ph/9906542}].

\bibitem{NO}    S. Nojiri and S.D. Odintsov, \eprint{hep-th/0409244}.

\bibitem{strif} S. Kachru, R. Kallosh, A. Linde, and S.P. Trivedi, Phys. Rev. D \textbf{68}, 046005 (2003) [\eprint{hep-th/0301240}].

							  S. Kachru, R. Kallosh, A. Linde, J. Maldacena, L. McAllister, and S.P. Trivedi, J. Cosmol. Astropart. Phys. 10 (2003) 013 [\eprint{hep-th/0308055}].

								N. Iizuka and S.P. Trivedi, Phys. Rev. D \textbf{70}, 043519 (2004) [\eprint{hep-th/0403203}].
								
\bibitem{ABF1}  G. Allemandi, A. Borowiec, and M. Francaviglia, Phys. Rev. D \textbf{70}, 043524 (2004) [\eprint{hep-th/0403264}].								
								
\bibitem{AM}    K. Aoyanagi and K. Maeda, Phys. Rev. D \textbf{70}, 123506 (2004) [\eprint{hep-th/0408008}].

\bibitem{crit}  G.W. Gibbons, \eprint{hep-th/0302199}.

							  J.M. Cline, S. Jeon, and G.D. Moore, Phys. Rev. D \textbf{70}, 043543 (2004) [\eprint{hep-ph/0311312}].

								S.M. Carroll, A. De Felice, and M. Trodden, \eprint{astro-ph/0408081}.
								
								C. Cs\'{a}ki, N. Kaloper, and J. Terning, \eprint{astro-ph/0409596}.
								
\bibitem{HN}    V.K. Onemli and R.P. Woodard, Class. Quantum Grav. \textbf{19}, 4607 (2002) [\eprint{gr-qc/0204065}].

								V.K. Onemli and R.P. Woodard, Phys. Rev. D \textbf{70}, 107301 (2004) [\eprint{gr-qc/0406098}].

								R. Holman and S. Naidu, \eprint{astro-ph/0408102}.

								A. Lue and G.D. Starkman, Phys. Rev. D \textbf{70}, 101501 (2004) [\eprint{astro-ph/0408246}].

\bibitem{fra02} P.H. Frampton, Phys. Lett. B \textbf{555}, 139 (2003) [\eprint{astro-ph/0209037}].

\bibitem{BDE}   E. Babichev, V. Dokuchaev, and Yu. Eroshenko, Phys. Rev. Lett. \textbf{93}, 021102 (2004)
[\eprint{gr-qc/0402089}].
 
 							  P.F. Gonz\'{a}lez-D\'{\i}az, Phys. Rev. Lett. \textbf{93}, 071301 (2004) [\eprint{astro-ph/0404045}].
							  
\bibitem{ban04} T. Banks, talk given at Cargèse Summer School, Cargèse (2004).

\bibitem{calv1} G. Calcagni, \eprint{gr-qc/0410027} (v1).

\end{thebibliography}
\end{document}